\documentclass[11pt]{article}
\usepackage[margin=1in]{geometry}
\usepackage{booktabs}
\usepackage{adjustbox}
\usepackage{graphicx}
\usepackage{amsmath}
\usepackage{subcaption}
\usepackage{setspace}
\usepackage{titlesec}
\usepackage{graphicx}
\usepackage[
    style=numeric,
    sorting=none,
    backend=biber,
    natbib=true 
]{biblatex}

\titleformat{\section}{\large\bfseries}{\thesection}{1em}{}
\titleformat{\subsection}{\normalsize\bfseries}{\thesubsection}{1em}{}

\title{\Large\textbf{Discrete-Time Survival Analysis for Heart Failure Mortality Prediction}}

\title{\Large\textbf{Discrete-Time Survival Analysis for Heart Failure Mortality Prediction}}

\author{
Aditya Rane$^{1*}$,
Amit Choudhari$^{2}$,
Shashi Kant$^{3}$,
and Akash Deep$^{1*}$\\[5pt]
\small $^{1}$Oklahoma State University, Stillwater, Oklahoma, USA\\
\small $^{2}$Cleveland State University, Cleveland, Ohio, USA\\
\small $^{3}$Cleveland Clinic, Cleveland, Ohio, USA\\[3pt]
\small Email:
\texttt{aditya.rane10@okstate.edu};
\texttt{akash.deep@okstate.edu}
}

\date{}
\begin{document}

\maketitle

\begin{abstract}
Accurate heart-failure prognosis relies on tracking clinical risk over time,
yet many machine-learning applications mishandle right-censored survival data
by either discarding a patient's observation time or using it as a predictor.
Discarding time ignores survival context, while using follow-up time as an
input feature introduces severe target leakage that inflates apparent accuracy.
We address this by proposing a discrete-time person-period framework for
heart-failure mortality classification. Using the UCI Heart Failure Clinical
Records cohort ($n=299$, 96 deaths), we transform the data into interval-level
binary outcomes and benchmark a Cox proportional hazards baseline against
person-period complementary log-log GLM and GAM models, alongside person-period
random forest, XGBoost, random survival forest, and DeepSurv classifiers. The
person-period GLM reproduces the Cox hazard ratios and concordance, validating
the transformation, while the GAM captures significant nonlinear predictor
effects and provides the best balance of discrimination and generalization; the
flexible classifiers achieve strong raw performance but overfit. Finally, we
quantify the leakage effect directly, including observed follow-up duration
raises classification AUC from roughly 0.73 to nearly 1.00, confirming that
follow-up duration must not be used as a baseline predictor. Overall, these
results establish a survival-aware framework that combines flexible
classification with valid time-to-event structure.
\end{abstract}

\section{Introduction} 

Heart failure is a chronic clinical condition associated with substantial
mortality, and accurate prognosis is central to patient management. Reliable risk prediction can help
practitioners identify high-risk patients, prioritize monitoring, and support
timely clinical decision making. Because the risk of death evolves throughout
follow-up, heart-failure prognosis is fundamentally a time-to-event problem:
each patient is observed for a period during which death either occurs or the
patient remains event-free until the last recorded contact.

Despite this time-to-event structure, much of the machine-learning (ML)
literature on heart-failure mortality reduces the problem to static binary
classification. One line of work discards follow-up time and labels each
patient according to whether death was observed
\citep{badik2024machine,chicco2020machine,zaman2021survival}. This formulation
does not fully account for censoring and treats early and late deaths as
equivalent outcomes, even though they contain different survival information.
A second approach includes observed follow-up duration as an ordinary input
feature. However, follow-up time is recorded after enrollment and is jointly
determined by the event and censoring processes. It would therefore not be
available when making a true baseline prediction. Its inclusion can introduce
outcome-dependent temporal information leakage and substantially inflate
apparent predictive performance without representing genuine baseline
clinical risk.

\begin{figure}[ht]
    \centering
    \includegraphics[width=\linewidth]{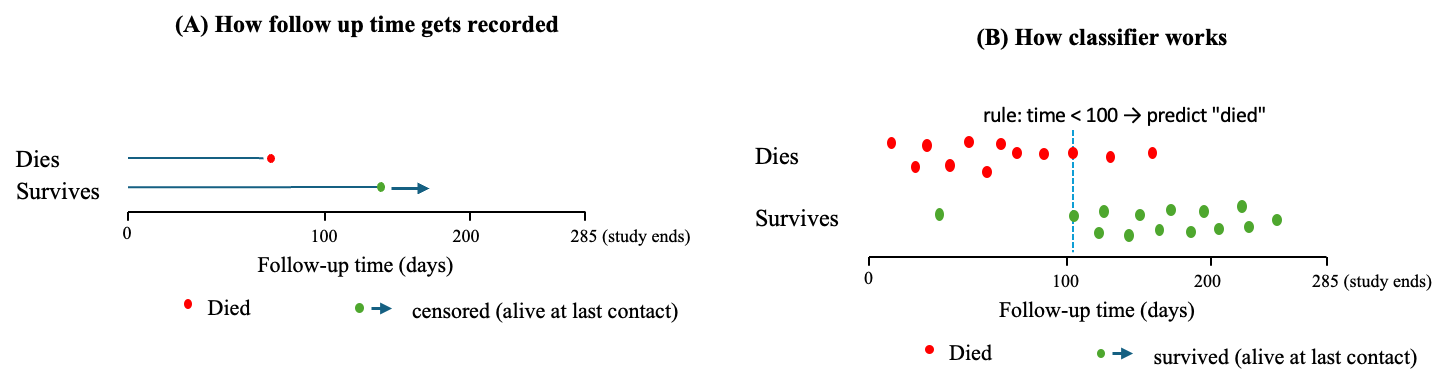}
    \caption{Illustration of observed follow-up duration and
    outcome-dependent information leakage in heart-failure mortality
    prediction.}
    \label{fig:illus1}
\end{figure}

The Heart Failure Clinical Records dataset contains both a death-event
indicator and an observed follow-up duration for each patient, making the
outcome inherently time-to-event rather than a simple yes/no label. The
primary prediction horizon in this study is prespecified at 180 days,
providing a common time point within the observed follow-up range for
evaluating baseline mortality risk. Figure~\ref{fig:illus1} illustrates why
observed follow-up duration cannot be treated as an ordinary baseline
covariate. On the left, patients have different observed follow-up durations:
some experience death during follow-up, whereas others are right-censored at
their last recorded contact. On the right, once a fixed prediction horizon is
defined, observed follow-up duration becomes strongly connected to the
resulting outcome label. Supplying it as an input feature can therefore reveal
outcome-related information and artificially inflate classification
performance.

To address this problem, we adopt a discrete-time person-period formulation of
the survival outcome. Cox proportional hazards regression provides the
continuous-time reference model \citep{cox1972regression}, after which the data
are transformed into a person-period structure and analyzed using a
complementary log-log model \citep{suresh2022survival}. This transformation
recasts survival analysis as an interval-level binary prediction problem while
retaining event timing and censoring. It therefore allows conventional
classification algorithms to be applied without using observed follow-up
duration as a baseline predictor.

Within this framework, a person-period generalized linear model (GLM) is used
to evaluate agreement with the Cox proportional hazards model. A generalized
additive model (GAM) is then used to examine nonlinear associations between
clinical predictors and mortality risk
\citep{hastie1986generalized,nelder1972generalized}. Flexible person-period
classification models, including Random Forest and XGBoost, are also
evaluated. In addition, Random Survival Forest and DeepSurv are included as
survival-aware machine-learning comparators.

The overall goal of this study is to establish a survival-aware evaluation
framework that connects traditional survival analysis with flexible
classification methods while preserving the time-to-event structure of the
data. The main contributions of this study are as follows:

\begin{itemize}
    \item Quantification of the apparent performance inflation caused by
    including outcome-dependent observed follow-up duration as a baseline
    predictor.

    \item Evaluation of a person-period framework that preserves event timing
    and censoring while enabling the use of conventional classification
    algorithms.

    \item Comparison of linear, nonlinear, and machine-learning survival
    models using repeated patient-level validation, with assessments of
    discrimination, prediction error, and overfitting.
\end{itemize}

\section{Overview of the Paper}
\label{sec:overview}

The remainder of this paper is organized as follows.
Section~\ref{sec:related_work} reviews prior work on heart-failure survival
analysis, machine-learning mortality classification, and discrete-time
survival modeling. Section~\ref{sec:data} describes the dataset and exploratory
analysis. Section~\ref{sec:methodology} presents the Cox baseline, the
person-period transformation, the statistical and machine-learning survival
models, the validation strategy, and the evaluation metrics.
Section~\ref{sec:results} reports predictive performance, nonlinear predictor
effects, and the impact of including observed follow-up duration.
Section~\ref{sec:conclusion} concludes and outlines future work.

\section{Related Work}
\label{sec:related_work}

Heart-failure mortality prediction has been studied using traditional 
survival-analysis methods, machine-learning (ML) classification algorithms, 
and more recent survival-aware ML approaches. The heart-failure clinical 
records dataset was originally analyzed as a time-to-event dataset in which 
each patient has an observed follow-up duration and a death-event indicator. 
Accordingly, prior research is organized into three areas: classical survival 
analysis, ML-based mortality classification, and approaches that reformulate 
survival prediction as a classification problem while preserving the 
time-to-event structure.

\subsection{Survival Analysis Approaches}

Traditional survival-analysis methods, including Kaplan--Meier estimation and 
Cox proportional hazards regression, have been used to investigate mortality 
among patients with heart failure \cite{ahmad2017survival}. Kaplan--Meier 
analysis provides a nonparametric description of the survival distribution, 
whereas Cox proportional hazards regression estimates the association between 
baseline clinical predictors and the hazard of death. Subsequent work on the 
same dataset developed gender-specific survival models to examine whether 
mortality patterns and predictor effects differed between male and female 
patients \cite{zahid2019gender}.

These studies appropriately incorporate follow-up time and censoring into the 
outcome definition. However, conventional Cox proportional hazards models 
generally represent continuous covariate effects linearly on the log-hazard 
scale and rely on the proportional hazards assumption. Consequently, they may 
not fully capture nonlinear predictor effects or complex relationships among 
clinical variables. This motivates the evaluation of more flexible statistical 
and machine-learning models that retain the time-to-event structure of the data.

\subsection{Machine-Learning Classification Approaches}

Recent advancements motivated applications of ML algorithms to the heart-failure clinical 
records dataset by treating death-event status as a binary classification 
outcome. Chicco and Jurman compared several classification methods and 
identified serum creatinine and ejection fraction as particularly important 
predictors of mortality \cite{chicco2020machine}. Other studies have evaluated 
logistic regression, support vector machines, random forests, gradient 
boosting, XGBoost, CatBoost, ensemble methods, and neural networks for 
predicting death-event status 
\cite{badik2024machine,moreno2023improvement,jia2024}. These studies demonstrate 
the potential of flexible ML algorithms for identifying complex relationships 
between clinical predictors and mortality.

However, reducing survival data to a single binary label can discard important 
time-to-event information. For example, patients who die early and patients who 
die late receive the same event label even though their survival experiences 
are substantially different. Similarly, a patient who is event-free after a 
short follow-up period does not provide the same information as a patient who 
remains event-free for a much longer period. Treating these observations as 
equivalent can result in an incomplete or misleading assessment of mortality 
risk.

A further concern arises when observed follow-up duration is included as an 
ordinary input feature in a mortality-classification model. Follow-up time is 
not a baseline clinical characteristic available at the time of prediction. 
Instead, it is observed after enrollment and represents either the time to 
death for a patient who experiences the event or the time to last contact for 
a censored patient. It is therefore directly related to the observed outcome. 
Using follow-up duration as a baseline predictor can introduce 
outcome-dependent information leakage and artificially inflate apparent 
predictive performance. Although previous studies have reported high 
classification accuracy on this dataset, the extent to which performance can 
be driven by follow-up information rather than genuine baseline clinical risk 
remains insufficiently examined.

A parallel line of research has developed ML methods that directly accommodate 
right-censored time-to-event outcomes. Random survival forests extend the 
random-forest framework by using survival-based node-splitting rules and 
estimating cumulative hazard or survival functions within terminal nodes 
\cite{Ishwaran_2008}. DeepSurv uses a neural-network representation of the 
Cox proportional hazards model and optimizes the Cox partial likelihood, 
allowing nonlinear and interaction effects to be learned while retaining a 
survival-analysis objective \cite{katzmandeepsurv}. These methods show that 
flexible ML models can be adapted to censored survival outcomes without 
reducing the problem to ordinary binary classification.

\subsection{Reformulating Survival Analysis as Classification}

Recent research has also demonstrated that survival analysis can be 
reformulated as a classification problem when event timing, risk sets, and 
censoring are appropriately retained. Craig et al.\ proposed ``survival 
stacking,'' which restructures right-censored data into risk-set-based binary 
outcomes so that standard classification algorithms can be applied within a 
survival setting \cite{craig2021survival}. This formulation is closely related 
to the Cox partial likelihood and illustrates that time should be used to 
construct valid risk sets rather than supplied to a model as a baseline 
predictor.

A related approach is discrete-time person-period modeling. In this framework, 
the follow-up period is divided into prespecified intervals, and each patient 
contributes one record for every interval during which the patient remains at 
risk. The interval-level binary outcome indicates whether the event occurred 
during that interval. Suresh et al.\ demonstrated how traditional regression 
and machine-learning classification algorithms can be fitted to person-period 
data and how interval-specific event probabilities can be combined to obtain 
patient-level survival probabilities \cite{suresh2022survival}. This framework 
allows flexible binary classifiers to be used while preserving event timing 
and appropriately incorporating right-censored observations.

Motivated by prior work, the present study focuses on the methodological 
consequences of using observed follow-up duration as an ordinary baseline 
predictor. The study systematically quantifies the resulting performance 
inflation and evaluates a person-period framework as a survival-aware 
alternative. It further compares linear, nonlinear, and machine-learning 
survival models under repeated patient-level validation, with explicit 
assessment of discrimination, prediction error, and overfitting.

\section{Data Description and Exploratory Data Analysis}
\label{sec:data}

\subsection{Data Description}

This study uses the Heart Failure Clinical Records dataset, which contains
medical records collected from patients treated at the Institute of Cardiology
and Allied Hospital in Pakistan between April and December 2015
\cite{ahmad2017data}. The dataset includes 299 patients with heart failure,
comprising 194 men and 105 women, all of whom were at least 40 years of age.
The patients had left ventricular systolic dysfunction and were classified as
having advanced-stage heart failure.

For the present analysis, the variables were categorized as baseline clinical
predictors or survival outcomes to ensure an appropriate time-to-event
formulation. The baseline predictors include demographic characteristics,
clinical conditions, lifestyle factors, and laboratory measurements, as
summarized in Table~\ref{tab:features}. The survival outcome is defined by the
observed follow-up duration and the death-event indicator.

\begin{table}[ht]
    \centering
    \caption{Clinical features and survival outcomes
    ($n=299$; 96 deaths and 203 right-censored observations).}
    \label{tab:features}
    \renewcommand{\arraystretch}{1.15}
    \begin{tabular}{p{3.8cm} p{1.8cm} p{7.5cm}}
        \hline
        \textbf{Feature} & \textbf{Type} & \textbf{Description} \\
        \hline
        \multicolumn{3}{c}{\textit{Baseline clinical predictors}} \\
        \hline
        Age & Numeric & Age in years \\
        Sex & Binary & Biological sex (male/female) \\
        Anaemia & Binary & Reduced red blood cell count \\
        Diabetes & Binary & Diabetes diagnosis \\
        High blood pressure & Binary & Hypertension \\
        Smoking & Binary & Active smoking status \\
        Ejection fraction & Numeric & Percentage of blood expelled per contraction \\
        Serum creatinine & Numeric & Serum creatinine level (mg/dL), a marker of renal function \\
        Serum sodium & Numeric & Serum sodium level (mEq/L) \\
        Platelets & Numeric & Platelet count (kilo/$\mu$L) \\
        Creatinine phosphokinase & Numeric & CPK enzyme level (mcg/L) \\
        \hline
        \multicolumn{3}{c}{\textit{Survival outcomes}} \\
        \hline
        Time (follow-up) & Numeric & Observed follow-up duration in days \\
        Death event & Binary & Death observed during follow-up (1) or right-censored (0) \\
        \hline
    \end{tabular}
\end{table}

\subsection{Exploratory Data Analysis}

Tables~\ref{tab:cont-summary} and~\ref{tab:bin-summary} summarize the continuous
and binary variables, respectively, for the 299 patients in the cohort.

As shown in Table~\ref{tab:cont-summary}, the continuous predictors differ
substantially in scale and distribution. Serum creatinine has a mean of
1.39 mg/dL and a median of 1.10 mg/dL. In contrast, creatinine phosphokinase
(CPK) is strongly right-skewed, with a mean of 581.84 mcg/L, a median of
250.0 mcg/L, and a maximum value of 7861.0 mcg/L. Platelet count is measured
on a substantially larger numerical scale than the other clinical variables.

These differences motivate model-specific preprocessing and the use of
methods capable of representing nonlinear predictor effects. Standardization
is applied where required for numerical stability, particularly for
neural-network models, whereas tree-based models do not require predictors
to be placed on a common scale.

\begin{table}[ht]
    \centering
    \caption{Summary statistics for continuous variables ($n=299$).}
    \label{tab:cont-summary}
    \renewcommand{\arraystretch}{1.15}
    \begin{tabular}{l r r r r r}
        \hline
        \textbf{Feature} & \textbf{Mean} & \textbf{SD} &
        \textbf{Median} & \textbf{Min} & \textbf{Max} \\
        \hline
        Age (years)                       &  60.83 &  11.89 &  60.0 &  40.0 &   95.0 \\
        Creatinine phosphokinase (mcg/L) & 581.84 & 970.29 & 250.0 &  23.0 & 7861.0 \\
        Ejection fraction (\%)            &  38.08 &  11.83 &  38.0 &  14.0 &   80.0 \\
        Platelets (kilo/$\mu$L)           & 263.36 &  97.80 & 262.0 &  25.1 &  850.0 \\
        Serum creatinine (mg/dL)          &   1.39 &   1.03 &   1.1 &   0.5 &    9.4 \\
        Serum sodium (mEq/L)              & 136.63 &   4.41 & 137.0 & 113.0 &  148.0 \\
        Time (days)                       & 130.26 &  77.61 & 115.0 &   4.0 &  285.0 \\
        \hline
    \end{tabular}
\end{table}

Table~\ref{tab:bin-summary} presents the frequency distributions of the binary
clinical predictors and the event indicator. Men represented 64.88\% of the
cohort. Anaemia was present in 43.14\% of patients, diabetes in 41.81\%, high
blood pressure in 35.12\%, and active smoking in 32.11\%.

\begin{table}[ht]
    \centering
    \caption{Frequency summary for binary variables ($n=299$).}
    \label{tab:bin-summary}
    \renewcommand{\arraystretch}{1.15}
    \begin{tabular}{l r r}
        \hline
        \textbf{Feature} & \textbf{$n$ (= 1)} & \textbf{Percent} \\
        \hline
        Death event (died)  &  96 & 32.11 \\
        Anaemia             & 129 & 43.14 \\
        Diabetes            & 125 & 41.81 \\
        High blood pressure & 105 & 35.12 \\
        Sex (male)          & 194 & 64.88 \\
        Smoking             &  96 & 32.11 \\
        \hline
    \end{tabular}
\end{table}

\begin{figure}[ht]
    \centering
    \includegraphics[width=0.85\linewidth]{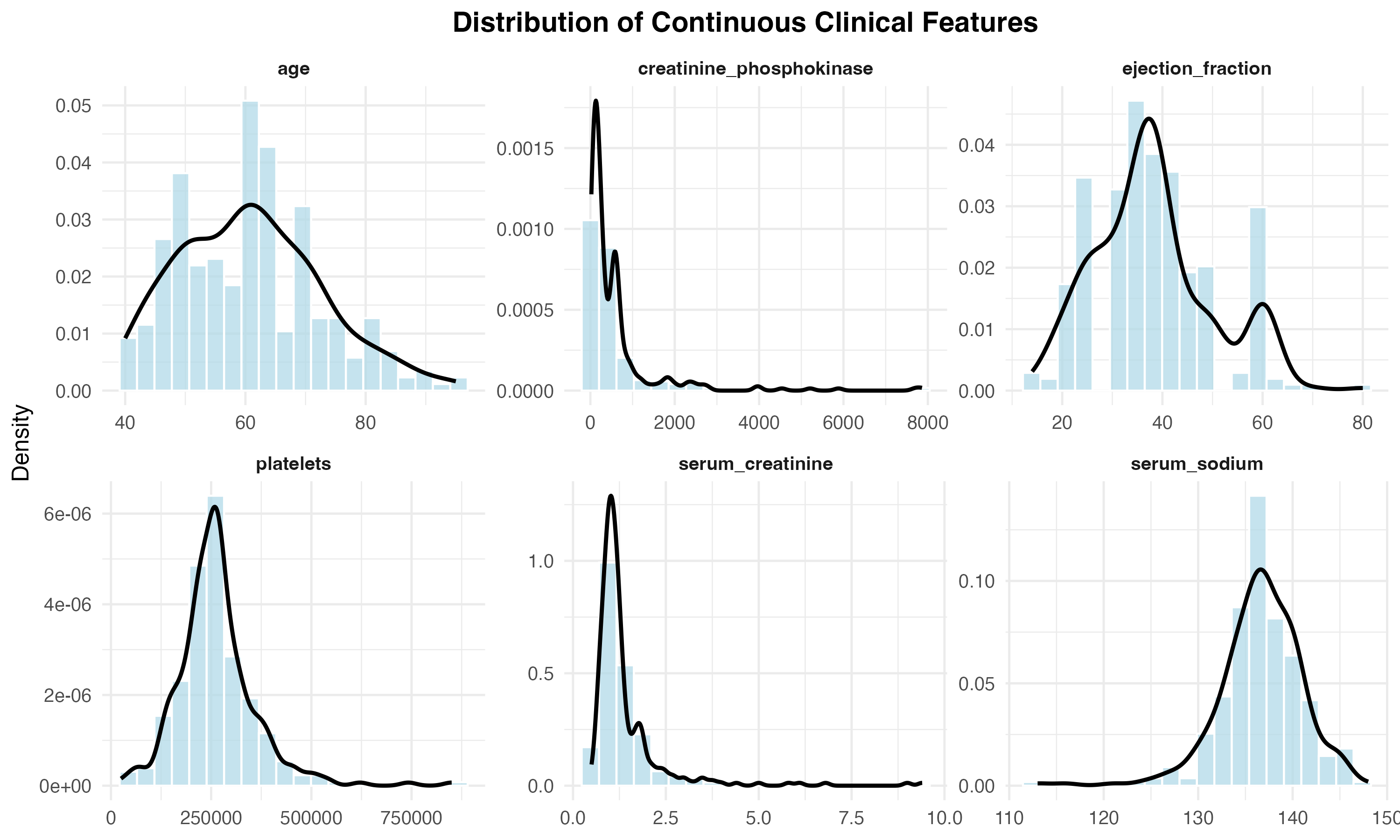}
    \caption{Distributions of the continuous clinical predictors.
    Histograms and density curves illustrate differences in scale, spread,
    and skewness across variables.}
    \label{fig:feature_dist}
\end{figure}

Figure~\ref{fig:feature_dist} displays the distributions of the continuous
clinical predictors. Creatinine phosphokinase and serum creatinine are
strongly right-skewed, with most observations concentrated at lower values
and a small number of observations extending into the right tails. Serum
sodium is concentrated within a relatively narrow range, whereas ejection
fraction has a non-uniform distribution. These patterns suggest that strictly
linear predictor effects may not adequately describe all relationships with
mortality risk and motivate the evaluation of generalized additive models.

Death was observed in 96 patients, corresponding to 32.11\% of the cohort.
The remaining 203 patients had no observed death and were right-censored at
their last recorded follow-up. Follow-up duration ranged from 4 to 285 days,
with a median of 115 days. The combination of variable follow-up durations,
observed deaths, and right-censored observations confirms that the outcome has
a time-to-event structure rather than a simple binary structure.

The dataset also presents two important modeling challenges. First, the sample
contains only 299 patients and 96 observed deaths. This limited effective
sample size increases uncertainty in estimated predictor effects and creates a
substantial risk of overfitting, particularly for flexible machine-learning
and deep-learning models. Second, deaths represent 32.11\% of the cohort,
compared with 67.89\% without an observed death. Although this represents a
moderate rather than extreme class imbalance, it may still affect
threshold-dependent measures such as accuracy, sensitivity, precision, and
F1 score. These characteristics motivate the use of parsimonious models,
patient-level repeated validation, censoring-aware performance measures, and
explicit assessment of training--test performance gaps.

\section{Methodology}
\begin{figure}
    \centering
    \includegraphics[width=1\linewidth]{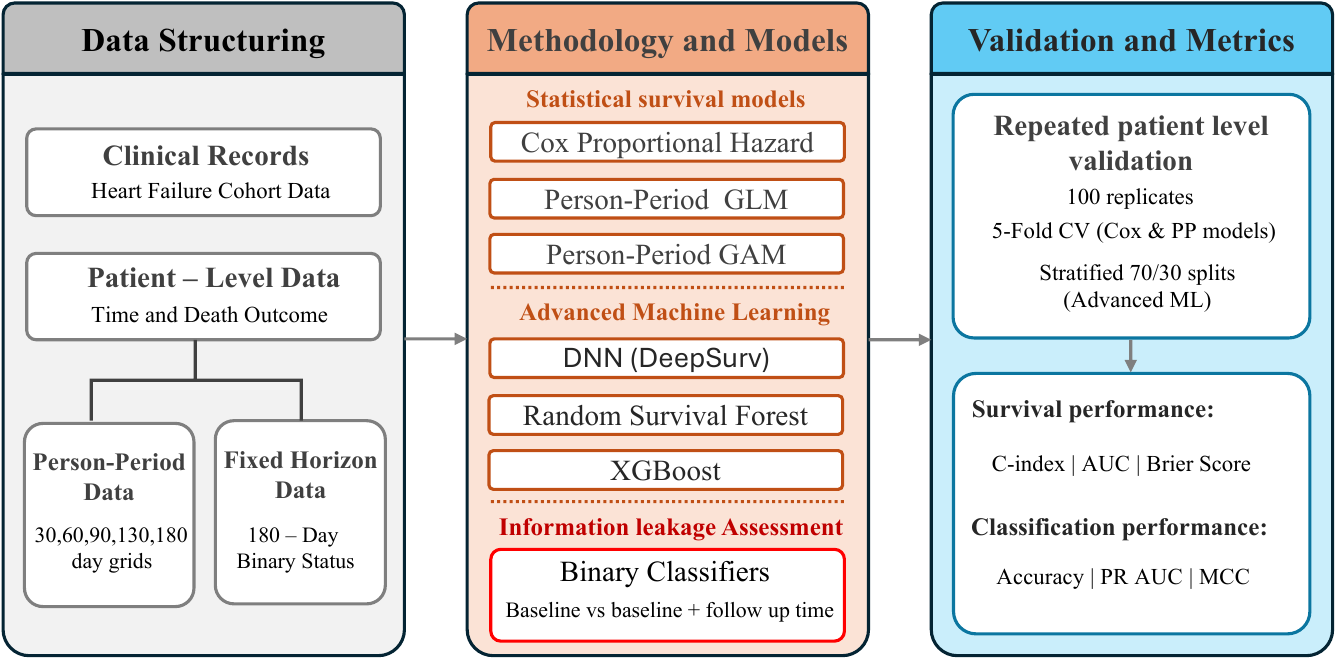}
    \caption{Methodological Framework}
    \label{fig:method_framework}
\end{figure}

Figure~\ref{fig:method_framework} summarizes the overall analytical framework.
The original patient-level data were analyzed through two parts. First, the survival-modeling retained all patients and incorporated observed follow-up duration and right censoring directly. Second, the fixed-horizon
classification pathway defined mortality status at 180 days and included only
patients whose outcome status was known at that horizon. Model development and
evaluation were performed using patient-level repeated validation to prevent
records from the same patient from appearing in both the training and test
sets.
\label{sec:methodology}
\subsection{Analysis Overview and Prediction Tasks}
\label{sec:analytical_framework}
This study consists of two related prediction analyses. The first evaluates
mortality using survival models that retain each patient's observed follow-up
duration and censoring indicator. The second evaluates mortality as an
ordinary binary classification problem at a common prediction horizon in order
to quantify the apparent performance gain produced by including observed
follow-up duration as an input feature.

The primary prediction horizon was prespecified at 180 days. For the
fixed-horizon classification analysis, a patient was classified as having an
event if death occurred on or before 180 days. A patient was classified as
event-free at 180 days if the patient remained under observation through at
least 180 days without an earlier death. Deaths occurring after 180 days were
therefore treated as event-free at the 180-day horizon.

Patients who were right-censored before 180 days had an unknown 180-day
mortality status. These patients were excluded only from the ordinary
fixed-horizon classification analysis because assigning them to the event-free
class would incorrectly assume that they survived through the complete
prediction horizon.

In contrast, all 299 patients were retained in the survival-model analyses.
For these analyses, the observed follow-up duration and death-event indicator
were jointly represented using a right-censored survival outcome. Patients
without an observed death contributed survival information through their last
recorded follow-up time. Censoring was accommodated directly during model
estimation and through censoring-aware performance measures, including
time-dependent area under the receiver operating characteristic curve and the
inverse-probability-of-censoring-weighted Brier score.

This separation ensures that patients with unknown 180-day status are not
incorrectly labeled in the ordinary classification analysis while their
available follow-up information remains fully utilized in the survival
analysis.

\subsection{Prediction Horizon and Interval Definitions}
\label{sec:intervals}

The primary prediction horizon was set at 180 days to provide a common time
point for evaluating mortality risk within the observed follow-up period. For
the discrete-time survival analyses, follow-up was partitioned at points
at 30, 60, 90, 130, and 180 days. These points defined the intervals
$(0,30]$, $(30,60]$, $(60,90]$, $(90,130]$, $(130,180]$, and
$(180,\infty)$ days.

Each patient contributed one person-period record for every interval in which
the patient remained at risk. For patients who experienced death, the
interval-level event indicator was coded as one in the interval containing the
death time and zero in all preceding intervals. Right-censored patients
contributed event-free records through their last observed follow-up interval.

For each patient, the predicted conditional event probabilities from the
intervals ending on or before 180 days were combined to obtain the cumulative
180-day mortality risk. This formulation preserves the ordering of event times
and the available follow-up information while allowing mortality to be modeled
as an interval-level binary outcome.

\subsection{Cox Proportional Hazards Model}
\label{sec:cox_model}

The Cox proportional hazards model was used as the continuous-time survival
baseline because it accounts for right-censored observations and provides
interpretable estimates of the relationships between baseline clinical
predictors and mortality risk \citep{cox1972regression}. For patient $i$, the
hazard of death at time $t$ was defined as

\begin{equation}
    h_i(t)
    =
    h_0(t)
    \exp\left(
        \beta_1 x_{i1}
        + \beta_2 x_{i2}
        + \cdots
        + \beta_p x_{ip}
    \right),
\end{equation}

where $h_0(t)$ represents the baseline hazard, $x_{ij}$ represents the value
of predictor $j$ for patient $i$, and $\beta_j$ represents the estimated
effect of that predictor.

The exponential of each estimated coefficient, $\exp(\beta_j)$, was
interpreted as a hazard ratio. A hazard ratio greater than one indicates that
higher values of the predictor are associated with a higher mortality hazard,
whereas a hazard ratio below one indicates a lower mortality hazard, holding
the remaining predictors constant.

The model was fitted using the observed follow-up duration and death-event
indicator for all patients. Patients who did not experience death during the
observed follow-up period were treated as right-censored and contributed
information through their last recorded follow-up time. Follow-up duration was
used only as part of the survival outcome and was not included as a baseline
clinical predictor.

The proportional hazards assumption was evaluated using scaled Schoenfeld
residuals. Predictor-specific and global tests were used to assess whether the
estimated predictor effects remained approximately constant over follow-up.

The probability of death by 180 days was calculated from the fitted Cox model
as

\begin{equation}
    \widehat{R}_i(180)
    =
    1 -
    \exp\left[
        -\widehat{H}_0(180)
        \exp\left(
            \widehat{\beta}_1 x_{i1}
            + \widehat{\beta}_2 x_{i2}
            + \cdots
            + \widehat{\beta}_p x_{ip}
        \right)
    \right],
\end{equation}

where $\widehat{H}_0(180)$ is the estimated cumulative baseline hazard at
180 days. Model performance was evaluated using the concordance index,
180-day time-dependent AUC, and inverse probability of censoring weighted Brier score.

\subsection{Person-Period Data Transformation}
\label{sec:pp_transformation}

The original dataset contained one patient-level record for each of the 299
patients, including baseline clinical predictors, observed follow-up duration,
and death-event status. To fit the discrete-time survival models, these data
were transformed into a person-period format. Under this representation, each
patient contributed one record for every follow-up interval during which the
patient remained under observation and at risk of death.

\begin{figure}[htbp]
    \centering
    \includegraphics[width=0.85\linewidth]{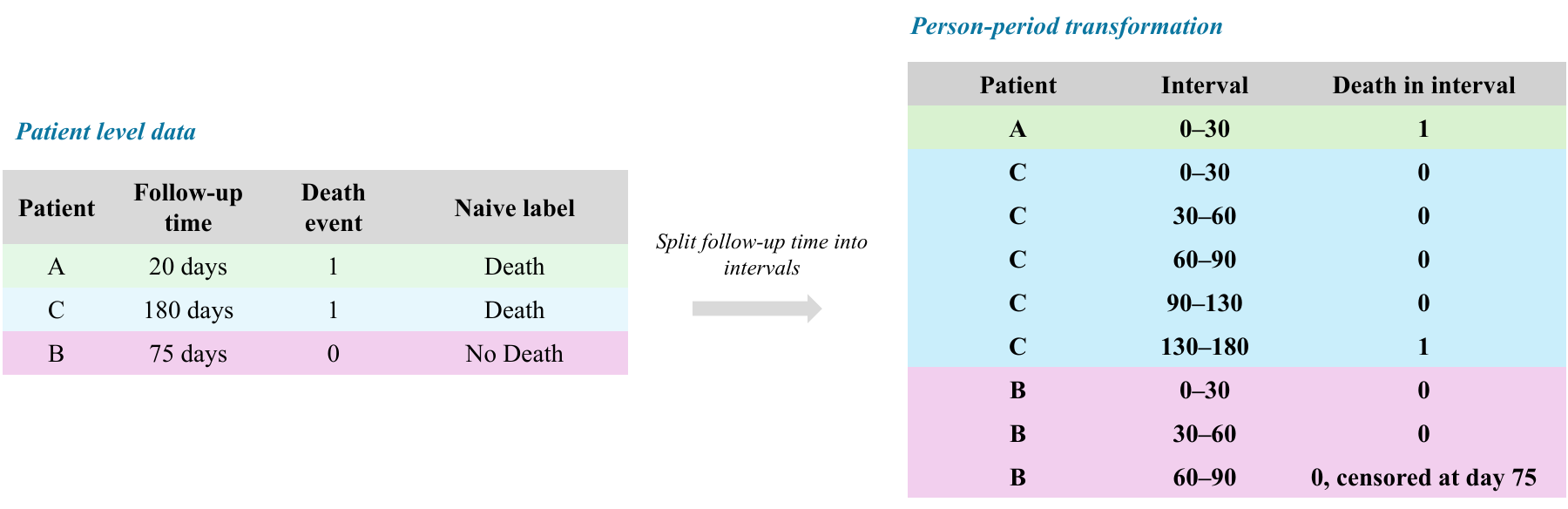}
    \caption{Illustration of person-period data transformation.}
    \label{fig:pp_transformation}
\end{figure}

Figure~\ref{fig:pp_transformation} illustrates the transformation from the
original patient-level structure to interval-level binary records. Patients
who experienced death contributed event-free records for the intervals before
death and an event record for the interval containing the death. Patients who
were right-censored contributed event-free records through the interval
containing their last observed follow-up time and no records after censoring.

Let the follow-up intervals be denoted by
$(t_{j-1},t_j]$, where $j=1,\ldots,J$. For patient $i$, the interval-specific
event indicator was defined as

\begin{equation}
    y_{ij}
    =
    \begin{cases}
        1, & \text{if patient $i$ died during interval $j$},\\
        0, & \text{otherwise}.
    \end{cases}
\end{equation}

Follow-up was divided into the intervals $(0,30]$, $(30,60]$, $(60,90]$,
$(90,130]$, $(130,180]$, and $(180,\infty)$ days. Each person-period record
contained the patient's baseline clinical predictors, an indicator for the
corresponding follow-up interval, and a binary event outcome indicating
whether death occurred during that interval.

For a patient who experienced death, the interval-level outcome was coded as
zero in all intervals before the death interval and as one in the interval
containing the death. No records were created after death. For a patient who
was censored, the interval-level outcome remained zero through the final
observed interval, and no records were created after the censoring time.
Observed follow-up duration was therefore used to determine the intervals
contributed by each patient rather than being included as a baseline
predictor.

The baseline clinical predictors were repeated across the interval-level
records belonging to the same patient. The interval indicator allowed the
baseline mortality risk to vary across follow-up intervals. Predictor effects
were otherwise assumed to remain constant across intervals unless nonlinear
or more flexible relationships were introduced by the fitted model.

The number of patients contributing records decreased over time as patients
experienced death or were censored. All 299 patients contributed to the first
interval, followed by 259 patients in the second interval, 236 in the third,
185 in the fourth, 133 in the fifth, and 103 in the final interval. As shown
in Table~\ref{tab:pp_summary}, the transformation expanded the dataset from
299 patient-level observations to 1,215 person-period observations while
retaining all 96 observed death events.

\begin{table}[htbp]
    \centering
    \caption{Person-period interval summary.}
    \label{tab:pp_summary}
    \renewcommand{\arraystretch}{1.15}
    \begin{tabular}{c c r}
        \hline
        \textbf{Interval} &
        \textbf{Follow-up interval} &
        \textbf{Patients at risk} \\
        \hline
        1 & 0--30 days          & 299 \\
        2 & 30--60 days         & 259 \\
        3 & 60--90 days         & 236 \\
        4 & 90--130 days        & 185 \\
        5 & 130--180 days       & 133 \\
        6 & More than 180 days  & 103 \\
        \hline
    \end{tabular}
\end{table}

For patient $i$, the conditional probability of death during interval $j$ was
defined as

\begin{equation}
    h_{ij}
    =
    \Pr\left(
        t_{j-1} < T_i \leq t_j
        \mid
        T_i > t_{j-1}
    \right),
    \label{eq:interval_hazard}
\end{equation}

where $T_i$ denotes the event time. Thus, $h_{ij}$ represents the probability
that patient $i$ experiences death during interval $j$, given that the patient
remained event-free at the beginning of that interval.

The predicted survival probability through interval $j$ was calculated by
multiplying the conditional probabilities of surviving each interval:

\begin{equation}
    S_i(t_j)
    =
    \prod_{k=1}^{j}
    \left(1-h_{ik}\right).
    \label{eq:discrete_survival}
\end{equation}

The corresponding cumulative mortality risk was calculated as

\begin{equation}
    R_i(t_j)
    =
    1-S_i(t_j).
    \label{eq:cumulative_risk}
\end{equation}

The predicted 180-day mortality risk was obtained by combining the
interval-specific event probabilities for the first five intervals, ending at
the 180-day horizon. The final interval, $(180,\infty)$, was retained for the
overall survival analysis but was not included in the calculation of 180-day
mortality risk. This transformation retains event timing and available
right-censoring information while allowing survival prediction to be
performed using interval-level binary classification models
\citep{suresh2022survival}.

\subsection{Person-Period Complementary Log-Log GLM}
\label{sec:pp_glm}

A generalized linear model with a complementary log-log link was fitted to
the person-period data. This model estimates the conditional probability that
a patient experiences death during a particular interval, given that the
patient remained at risk at the beginning of that interval. The complementary
log-log link was selected because it provides a discrete-time proportional
hazards formulation that can be directly compared with the continuous-time Cox
proportional hazards model \citep{suresh2022survival}.

For patient $i$ in interval $j$, let $p_{ij}$ denote the conditional probability
of death during that interval. The model was specified as

\begin{equation}
    \log\left[-\log\left(1-p_{ij}\right)\right]
    =
    \alpha_j
    +
    \beta_1 x_{i1}
    +
    \beta_2 x_{i2}
    +
    \cdots
    +
    \beta_p x_{ip},
    \label{eq:pp_cloglog}
\end{equation}

where $\alpha_j$ represents the effect of interval $j$, $x_{ik}$ represents
the value of clinical predictor $k$ for patient $i$, and $\beta_k$ represents
the estimated effect of that predictor. The interval indicator was entered as
a categorical variable, allowing the baseline mortality risk to vary across
follow-up intervals without imposing a fixed trend over time.
Thus, the model treats each person-period record as a binary prediction task:
whether the patient experienced death during that interval, conditional on
remaining at risk at the start of the interval.
The baseline clinical predictors were repeated across all person-period
records belonging to the same patient. In this model, the effects of the
clinical predictors were assumed to remain constant across intervals, while
the interval-specific terms allowed the underlying risk of death to change
over follow-up. Predictor-by-interval interactions were not included in the
primary model.

The exponential of each predictor coefficient, $\exp(\beta_k)$, was
interpreted as a hazard ratio. A value greater than one indicates that higher
values of the predictor are associated with an increased mortality hazard,
whereas a value below one indicates a reduced mortality hazard, holding the
remaining predictors constant.

The model was estimated using binomial maximum likelihood with a
complementary log-log link. Although each patient could contribute multiple
person-period records, the resulting binomial likelihood corresponds to the
likelihood of the discrete-time survival model. Thus, the repeated rows
represent the patient's event history across intervals rather than independent
patient-level outcomes \citep{suresh2022survival}.

The person-period complementary log-log GLM was used to assess whether the
discrete-time formulation reproduced the predictor effects and discrimination
of the Cox proportional hazards model. Interval-specific death probabilities
estimated by the model were combined, as described in
Subsection~\ref{sec:pp_transformation}, to obtain patient-level mortality risk at
the 180-day prediction horizon.

\subsection{Person-Period Generalized Additive Model}
\label{sec:pp_gam}

The person-period complementary log-log GLM assumes that each continuous
predictor has a linear association with the log cumulative interval hazard.
To allow more flexible relationships, a generalized additive model (GAM) was
fitted to the person-period data using the same complementary log-log link
\citep{hastie1986generalized, wood2017generalized}.

Age, ejection fraction, and serum creatinine were modeled using smooth
functions because these continuous predictors may have nonlinear associations
with mortality risk. The remaining clinical predictors were included as
linear terms. For patient $i$ in interval $j$, the model was specified as

\begin{equation}
\begin{split}
\log\left[-\log\left(1-p_{ij}\right)\right]
={}& \alpha_j
+ f_1(\text{age}_i)
+ f_2(\text{ejection fraction}_i) \\
&+ f_3(\text{serum creatinine}_i)
+ \sum_{r=1}^{q}\beta_r z_{ir},
\end{split}
\label{eq:pp_gam}
\end{equation}

where $p_{ij}$ is the conditional probability that patient $i$ experiences
death during interval $j$, given that the patient remained at risk at the
beginning of the interval. The term $\alpha_j$ represents the interval effect,
and $f_1$, $f_2$, and $f_3$ represent smooth functions for age, ejection
fraction, and serum creatinine, respectively. The terms $z_{ir}$ represent the
remaining clinical predictors, which were included linearly.

The interval indicator was entered as a categorical variable, allowing the
baseline mortality risk to differ across follow-up intervals. The clinical
predictor effects were assumed to remain constant across intervals; therefore,
predictor-by-interval interactions were not included in the primary model.

The smooth functions were estimated using penalized regression splines.
Candidate basis dimensions of $k=4,5,6,$ and $7$ were evaluated using
patient-level cross-validated predictive performance and model information
criteria. A basis dimension of $k=4$ was selected because larger values
provided only small improvements in predictive performance while increasing
model complexity. The selected value of $k=4$ was then fixed for all
subsequent repeated-validation analyses. Smoothing parameters were estimated
using restricted maximum likelihood.

Evidence of nonlinearity was assessed using the estimated degrees of freedom
(EDF), statistical significance of the smooth terms, and visual inspection of
the fitted smooth-effect plots. An EDF close to one indicates an approximately
linear relationship, whereas an EDF greater than one indicates increasing
evidence of a nonlinear relationship.

The fitted interval-specific event probabilities were combined across the
intervals ending on or before 180 days to obtain patient-level cumulative
mortality risk, as described in Section~\ref{sec:pp_transformation}. The
person-period GAM was then compared with the person-period GLM to determine
whether allowing nonlinear predictor effects improved mortality prediction
and model fit.

\subsection{Person-Period Machine-Learning Models}
\label{sec:pp_ml}

To capture nonlinear relationships and interactions that may not be fully
represented by the GLM or GAM, random forest and XGBoost classifiers were
applied to the person-period data. These models treated each person-period
record as a binary classification observation, with the outcome indicating
whether death occurred during the corresponding interval
\citep{suresh2022survival}.

For patient $i$ in interval $j$, the conditional probability of death was
represented generally as

\begin{equation}
    p_{ij}
    =
    f\left(
        x_{i1},x_{i2},\ldots,x_{ip},A_j
    \right),
    \label{eq:pp_ml}
\end{equation}

where $x_{i1},\ldots,x_{ip}$ denote the baseline clinical predictors,
$A_j$ denotes the categorical follow-up interval, and $f(\cdot)$ represents
the fitted model. Unlike the GLM, these models do not require
the clinical predictors to have linear or additive effects. They can therefore
capture nonlinear associations and interactions among the predictors and
follow-up intervals.

\subsubsection{Person-Period Random Forest}

The person-period random forest was fitted as a binary classifier using the
baseline clinical predictors and categorical interval indicator
\citep{breiman2001random}. Each tree was constructed from a bootstrap sample
of the person-period training data, and a random subset of predictors was
considered at each split. The predicted conditional probability of death in an
interval was obtained by averaging the class probabilities across all trees.

The final random forest contained 1,500 trees. The number of candidate
predictors considered at each split was set to the square root of the total
number of predictors. A minimum terminal-node size of 10 and a maximum of 30
terminal nodes per tree were used to limit model complexity. No additional
class weighting was applied in the final specification.

\subsubsection{Person-Period XGBoost}

XGBoost was also fitted to the person-period outcome using a binary logistic
objective \citep{chen2016xgboost}. The categorical interval indicator was
converted into binary indicator columns and included together with the
baseline clinical predictors. XGBoost constructs an ensemble of decision trees
sequentially, with each new tree attempting to reduce the prediction errors
remaining from the preceding trees.

The final model used 500 boosting rounds, a maximum tree depth of 3, and a
learning rate of 0.02. Regularization parameters were set to
$\lambda=5$ for the squared-coefficient penalty and $\alpha=0.50$ for the
absolute-coefficient penalty. A minimum loss reduction of $\gamma=0.50$ was
required for an additional tree split. Row and predictor subsampling rates
were both set to 0.80. The model was trained using binary log-loss, and no
additional positive-class weighting was applied in the final specification.

For prediction, each model estimated the conditional probability of death for
every interval ending on or before 180 days. The predicted interval-specific
probabilities were then converted into patient-level survival probability as

\begin{equation}
    S_i(180)
    =
    \prod_{j:t_j\leq180}
    \left(1-p_{ij}\right),
\end{equation}

and the corresponding 180-day mortality risk was calculated as

\begin{equation}
    R_i(180)
    =
    1-S_i(180).
\end{equation}

All training and testing splits were created at the patient level before the
person-period transformation. Therefore, interval records belonging to the
same patient could not appear in both the training and test sets. Model
performance and train--test differences were evaluated using the repeated
patient-level validation procedure described in
Section~\ref{sec:validation}.

\subsection{Continuous-Time Machine-Learning Survival Models}
\label{sec:continuous_ml}

Random survival forest and DeepSurv were included as flexible
continuous-time machine-learning survival models. Unlike the person-period
models, these approaches were fitted directly to the original patient-level
data using each patient's observed follow-up duration and death-event
indicator. Thus, all patients were retained, including those who were
right-censored before the 180-day prediction horizon. Observed follow-up
duration was used as part of the survival outcome and was not included as a
baseline predictor.

\subsubsection{Random Survival Forest}
Random survival forest extends the random forest algorithm to right-censored
time-to-event outcomes \citep{Ishwaran_2008}. Each survival tree was
constructed from a bootstrap sample of the training patients. At each node, a
random subset of predictors was considered, and the split producing the
greatest separation between the survival experiences of the resulting groups
was selected using a survival-based splitting rule.

A cumulative hazard function was estimated within each terminal node. For
patient $i$, the ensemble cumulative hazard was calculated by averaging the
cumulative hazard estimates from all trees:

\begin{equation}
    \widehat{H}_i(t)
    =
    \frac{1}{B}
    \sum_{b=1}^{B}
    \widehat{H}_{ib}(t),
    \label{eq:rsf_hazard}
\end{equation}

where $B$ is the total number of trees and
$\widehat{H}_{ib}(t)$ is the cumulative hazard estimate from tree $b$ for
patient $i$. The corresponding predicted survival probability was calculated
as

\begin{equation}
    \widehat{S}_i(t)
    =
    \exp\left[-\widehat{H}_i(t)\right],
    \label{eq:rsf_survival}
\end{equation}

and the predicted mortality risk by 180 days was obtained as

\begin{equation}
    \widehat{R}_i(180)
    =
    1-\widehat{S}_i(180).
    \label{eq:rsf_risk}
\end{equation}

The random survival forest was fitted using 500 trees, a minimum terminal-node
size of 60, and 10 candidate split points for each continuous predictor. The
default log-rank splitting rule was used. Both fitted training predictions and
out-of-bag predictions were retained to assess model generalization and
overfitting.

\subsubsection{DeepSurv}

DeepSurv is a neural-network extension of the Cox proportional hazards model
that replaces the linear predictor with a flexible nonlinear function of the
baseline clinical predictors \citep{katzmandeepsurv}. For patient $i$, the
hazard function was represented as

\begin{equation}
    h_i(t)
    =
    h_0(t)
    \exp\left[g(x_i)\right],
    \label{eq:deepsurv_hazard}
\end{equation}

where $h_0(t)$ is the baseline hazard and $g(x_i)$ is the risk score produced
by the neural network from the clinical predictors of patient $i$.

The neural network was trained by minimizing the negative Cox partial
log-likelihood:

\begin{equation}
    \mathcal{L}
    =
    -
    \sum_{i:\delta_i=1}
    \left[
        g(x_i)
        -
        \log
        \left\{
            \sum_{\ell \in \mathcal{R}_i}
            \exp\left[g(x_\ell)\right]
        \right\}
    \right],
    \label{eq:deepsurv_loss}
\end{equation}

where $\delta_i$ is the death-event indicator and $\mathcal{R}_i$ is the set
of patients still at risk immediately before the event time of patient $i$.
This loss function allows right-censored patients to contribute information
through the risk sets without treating censoring as a death event.

Continuous predictors were standardized using the means and standard
deviations estimated from the training data. The same training-based
standardization was then applied to the corresponding test data. The network
contained two hidden layers with 32 nodes in each layer, exponential linear
unit activation functions, and a single linear output node representing the
patient's relative risk score. Full-batch training was used with an
optimization parameter of $\alpha=0.50$.

Because neural-network estimation was sensitive to the small sample size,
predefined learning-rate, regularization, and epoch combinations were
evaluated using three random restarts with learning rates
$1\times10^{-6}$ and training with 1,500 epochs.

DeepSurv produces relative risk scores rather than direct mortality
probabilities. Therefore, the cumulative baseline hazard was estimated from
the training data using the Breslow estimator. This estimate was combined
with each patient's predicted risk score to obtain the corresponding
180-day mortality probability. The predicted mortality
risk for patient $i$ at 180 days was then calculated as

\begin{equation}
    \widehat{R}_i(180)
    =
    1-
    \exp\left[
        -\widehat{H}_0(180)
        \exp\left\{g(x_i)\right\}
    \right]
    \label{eq:deepsurv_risk}
\end{equation}

where $\widehat{H}_0(180)$ is the estimated cumulative baseline hazard at
180 days. Random survival forest and DeepSurv were evaluated using the same repeated
patient-level train--test splits as the other survival models. Their patient-level 180-day risk estimates were assessed using the  performance measures described in
Section~\ref{sec:performance}.

\subsection{Predictor Sets and Feature Selection}
The primary analysis used the complete set of 11 baseline clinical predictors
available in the dataset: age, anaemia, creatinine phosphokinase, diabetes,
ejection fraction, high blood pressure, platelet count, serum creatinine,
serum sodium, sex, and smoking status. This full predictor set was used as the
main analysis so that the models could be compared using the same baseline
clinical information without removing variables based on the observed
outcomes.

Observed follow-up duration and death-event status were not included as
baseline predictors. Follow-up duration and event status jointly defined the
survival outcome. For the person-period models, the follow-up duration was
used only to determine the intervals contributed by each patient. The interval
indicator was then added as a model term to represent changes in baseline risk
over time, but it was not treated as a candidate clinical predictor during
feature selection.

A fixed parsimonious predictor set was also evaluated as a sensitivity
analysis. This set contained age, anaemia, creatinine phosphokinase, ejection
fraction, high blood pressure, and serum creatinine. The parsimonious analysis
was used to examine whether reducing the number of predictors improved
generalization in the relatively small cohort. Results from the full predictor
set remained the primary findings, while results from the parsimonious set
were treated as supporting analyses. Furthermore, binary predictors were kept using their original indicator coding. The
continuous predictors were used on their original scales for the Cox,
person-period GLM, GAM, random forest, XGBoost, and random survival forest
models. For DeepSurv, continuous predictors were standardized using the mean
and standard deviation estimated from the training fold, and the same
training-based transformation was applied to the held-out fold.

\subsection{Validation Strategy}
\label{sec:validation}
Since an independent external validation cohort was not available, model
performance was evaluated using repeated patient-level internal validation.
Two validation procedures were used for different model comparison
objectives.

For the comparison of the statistical survival models performance, including the Cox
proportional hazards model, person-period complementary log-log GLM, and
person-period GAM, repeated five-fold patient-level cross-validation was used.
In each repetition, patients were divided into five mutually exclusive folds.
Four folds were used for model training, and the remaining fold was used for
testing. This process was repeated until each fold had served once as the test
set. The complete five-fold procedure was repeated 100 times, resulting into 500
held-out fold evaluations for each model.

All fold assignments were created at the patient level before the
person-period transformation. Within each fold, the training patients and test
patients were expanded into person-period records separately. Therefore,
multiple interval records belonging to the same patient could not appear in
both the training and test data. The same fold assignments were used for the
Cox PH, person-period GLM, and person-period GAM so that their performance was
compared using identical training and test patients.

Next, additional comparison to measure the classification performance of Cox PH, person-period GLM,
person-period GAM, person-period random forest, person-period XGBoost, random
survival forest, and DeepSurv, 100 repeated stratified 70/30 train--test
splits were used. In each repetition, 70\% of the patients were assigned to
the training set and 30\% to the test set. Stratification was based on the
death-event indicator so that the proportion of observed deaths remained
approximately similar in the training and test sets.

\subsection{Follow-Up-Time Information Assessment}
\label{sec:followup_assessment}
A separate fixed-horizon classification experiment was studied to assess
the effect of including observed follow-up duration as a predictor of
180-day mortality. This assessment was motivated by the fact that observed
follow-up duration is determined after baseline and is directly related to
outcome related information. Therefore, it is not a clinical
variable that would be available when making a baseline mortality prediction.

For this experiment, a binary 180-day outcome was defined. Patients who died
on or before 180 days were assigned an outcome of one. Patients known to be
event-free at 180 days were assigned an outcome of zero, including patients
whose death occurred after 180 days. Patients who were right-censored before
180 days were excluded from this fixed-horizon classification analysis because
their mortality status at the prediction horizon was unknown. These patients
remained included in all censoring-aware survival analyses.
Three predictor sets were evaluated:
\begin{enumerate}
    \item \textbf{Clinical predictors only:} the baseline clinical variables
    were used without observed follow-up duration.

    \item \textbf{Clinical predictors with follow-up duration:} observed
    follow-up duration was added to the baseline clinical variables. 

    \item \textbf{Follow-up duration only:} observed follow-up duration was
    used as the only predictor.
\end{enumerate}

The clinical-only setting represented a realistic baseline prediction task.
The clinical with follow-up setting measured the relative improvement
when outcome-dependent follow-up information was included. The follow-up only
setting examines how much of the 180-day outcome could be inferred from
follow-up duration without using any clinical information. Logistic regression, weighted logistic regression, random forest, and XGBoost
were evaluated under these three predictor settings. Weighted logistic
regression assigned greater importance to the less frequent outcome class, whereas the remaining models used their standard classification
specifications. Observed follow-up duration was not included in the
censoring-aware Cox, person-period, random survival forest, or DeepSurv models as a baseline predictor.

Similar validation strategy as discussed in Section \ref{sec:validation} are applied to fixed-horizon experiment. Within each repetition, the same training and test patients were used for all three predictor settings and all classifiers. This ensured that changes in predictive performance were caused by the information provided to the models rather than differences in the evaluated patient samples. Performance under the three predictor settings was compared using the
classification measures described in Section~\ref{sec:performance}. In particular, the difference between the clinical-only andclinical with follow-up settings was used to quantify the relative performance gain associated with including observed follow-up duration. 

\subsection{Performance Metrics}
\label{sec:performance}
Model performance was evaluated using censoring-aware survival metrics and
fixed-horizon classification metrics.

Survival performance was assessed using Harrell's concordance index,
180-day time-dependent AUC, and the inverse probability of censoring-weighted
Brier score. The concordance index measured how well the models ranked
patients according to mortality risk. The time-dependent AUC measured the
ability to distinguish patients who experienced death by 180 days from those
who remained event-free beyond 180 days while accounting for censoring \citep{heagerty2000time}. The
IPCW Brier score measured the accuracy of the predicted 180-day mortality
probabilities while adjusting for incomplete follow-up. Higher concordance
and AUC values indicate better discrimination, whereas lower Brier scores
indicate better probability prediction.

For the fixed-horizon analyses, performance was evaluated using ROC-AUC,
precision--recall AUC, accuracy, Matthews correlation coefficient, and the Brier score.
Binary predictions were obtained using a probability threshold of 0.50.
Higher values indicate better performance for all metrics except the Brier
score, for which lower values are preferred. All results were reported as mean $\pm$ one standard deviation across the
held-out validation evaluations.

\section{Results}
\label{sec:results}
\subsection{Comparison of Statistical Survival Models and Nonlinear Effects}
\label{sec:statistical_models}

\subsubsection{Cox PH and person-period GLM agreement.}

The Cox proportional hazards model was statistically significant overall
($p<0.001$) and produced a full-cohort C-index of 0.741, indicating moderate
discrimination. Six of the 11 baseline predictors showed evidence of an
association with mortality. Higher age, anaemia, high blood pressure, and
serum creatinine were associated with increased mortality hazard, whereas
higher ejection fraction was associated with reduced mortality hazard.
Creatinine phosphokinase also showed evidence of association, although its
hazard ratio rounded to 1.00 because the effect was estimated for a one-unit
increase.

The global Schoenfeld residual test did not indicate an overall violation of
the proportional-hazards assumption ($p=0.39$). However, ejection fraction
showed possible evidence of a time-varying effect ($p=0.03$). Cox PH was
therefore considered a reasonable statistical baseline, although the
association between ejection fraction and mortality may not have remained
constant throughout follow-up.

The person-period complementary log-log GLM produced hazard-ratio estimates
that were highly consistent with those from Cox PH. The estimated hazard
ratio for age was 1.05 in both models, while the estimate for ejection
fraction was 0.952 in both models. The hazard ratios for serum creatinine were
1.38 for Cox PH and 1.40 for the PP GLM. Similarly, the estimates for anaemia
were 1.58 and 1.52, while those for high blood pressure were 1.61 and 1.52,
respectively. The directions of association were consistent across all
predictors.

The agreement was also reflected in full-cohort discrimination. The C-index
was 0.7408 for Cox PH and 0.741 for the PP GLM. Thus, the person-period
transformation and complementary log-log link closely reproduced the
predictor relationships and risk ordering obtained from the continuous-time
Cox model.

\begin{table}[htbp]
    \centering
    \caption{Comparison of hazard-ratio estimates from Cox PH and the
    person-period complementary log-log GLM.}
    \label{tab:cox_pp_hr}
    \begin{tabular}{lcc}
        \hline
        \textbf{Predictor} &
        \textbf{Cox PH HR} &
        \textbf{PP GLM HR} \\
        \hline
        Age                      & 1.05  & 1.05  \\
        Anaemia                  & 1.58  & 1.52  \\
        Creatinine phosphokinase & 1.00  & 1.00  \\
        Diabetes                 & 1.15  & 1.22  \\
        Ejection fraction        & 0.952 & 0.952 \\
        High blood pressure      & 1.61  & 1.52  \\
        Platelets                & 1.00  & 1.00  \\
        Serum creatinine         & 1.38  & 1.40  \\
        Serum sodium             & 0.957 & 0.965 \\
        Sex                      & 0.789 & 0.793 \\
        Smoking                  & 1.14  & 1.12  \\
        \hline
    \end{tabular}
\end{table}

\subsubsection{Nonlinear effects identified by the Person period GAM.}

The PP GAM identified significant nonlinear associations between mortality
risk and age, ejection fraction, and serum creatinine. The estimated degrees
of freedom were 2.14 for age, 2.31 for ejection fraction, and 2.62 for serum
creatinine, with all three smooth terms significant at $p<0.001$. Because
each estimated degree of freedom was greater than one, these relationships
were more complex than simple linear effects.

The estimated smooth effect for age indicated that mortality risk increased
more strongly at older ages. Ejection fraction showed an overall inverse
association with mortality risk, although the strength of this relationship
varied across its observed range. Serum creatinine also showed a nonlinear
association, with mortality risk increasing more strongly at higher values.
The confidence bands widened near the extremes of the predictor ranges,
reflecting the smaller number of observations in those regions.

\begin{figure}[http]
    \centering
    \includegraphics[width=\linewidth]{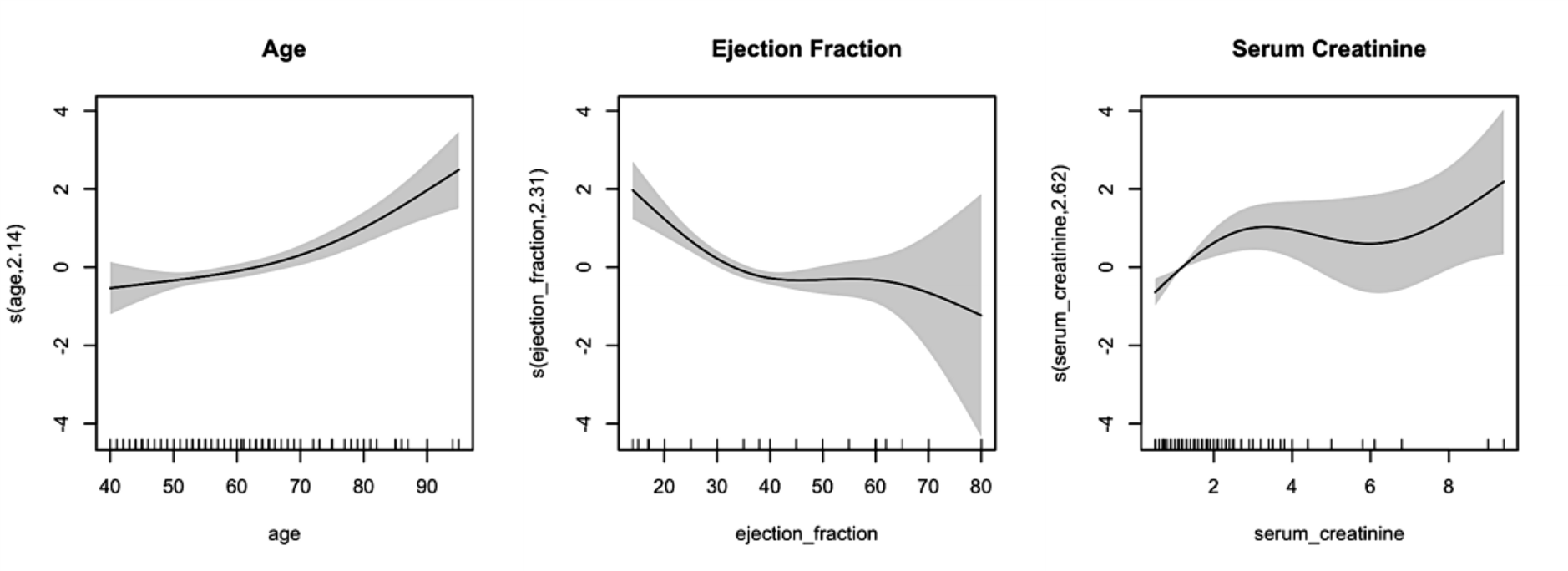}
    \caption{Estimated nonlinear effects of age, ejection fraction, and serum
    creatinine from the person-period GAM. Shaded regions represent
    uncertainty around the estimated smooth effects.}
    \label{fig:gam_smooth_effects}
\end{figure}

\subsubsection{GAM complexity selection.}

Candidate basis dimensions of $k=4,5,6,$ and $7$ resulted into similar
cross-validated discrimination. The mean AUC increased from 0.772 at $k=4$
to 0.780 at $k=5$ and 0.781 at both $k=6$ and $k=7$. Increasing the basis
dimension beyond four therefore yields only small improvements in
discrimination.

The model with $k=4$ had the lowest AIC and BIC values, at 486 and 592,
respectively. The corresponding BIC values increased from 598 to 602
for $k=5$, $k=6$, and $k=7$. Therefore, $k=4$ was preferred because it
provided a favorable balance between predictive performance and model
complexity.

\begin{table}[htbp]
    \centering
    \caption{Comparison of candidate basis dimensions for the person-period
    GAM.}
    \label{tab:gam_k_selection}
    \begin{tabular}{cccc}
        \hline
        \textbf{Basis dimension} &
        \textbf{Mean AUC} &
        \textbf{Mean AIC} &
        \textbf{Mean BIC} \\
        \hline
        $k=4$ & 0.772 & 486 & 592 \\
        $k=5$ & 0.780 & 487 & 598 \\
        $k=6$ & 0.781 & 488 & 601 \\
        $k=7$ & 0.781 & 488 & 602 \\
        \hline
    \end{tabular}
\end{table}

\subsubsection{Cross-validated performance using the full predictor set}

Repeated five-fold patient-level cross-validation showed that Cox PH and the
PP GLM had nearly identical held-out discrimination when all 11 baseline
predictors were used. Their held-out C-index values were approximately 0.71,
and their 180-day time-dependent AUC values were approximately 0.72. This
agreement indicates that the person-period complementary log-log model
preserved the predictive performance of the continuous-time Cox model.

The PP GAM achieved the highest held-out performance, with a C-index of
approximately 0.73 and an AUC of approximately 0.76. Although the improvement
was moderate relative to the variability across validation folds, both
discrimination measures favored the GAM, suggesting that nonlinear predictor
effects provided additional predictive information.

Across both person-period models, AIC favored the PP GAM, whereas BIC favored
the simpler PP GLM. This difference reflects the stronger complexity penalty
applied by BIC. The AIC and BIC values from Cox PH were not directly
compared with those from the person-period models because Cox PH uses a
partial likelihood on patient-level survival data, whereas the PP models use
a binomial likelihood on interval-level data.

\begin{figure*}[!t]
    \centering

    \begin{subfigure}[t]{0.32\textwidth}
        \centering
        \includegraphics[width=\linewidth]{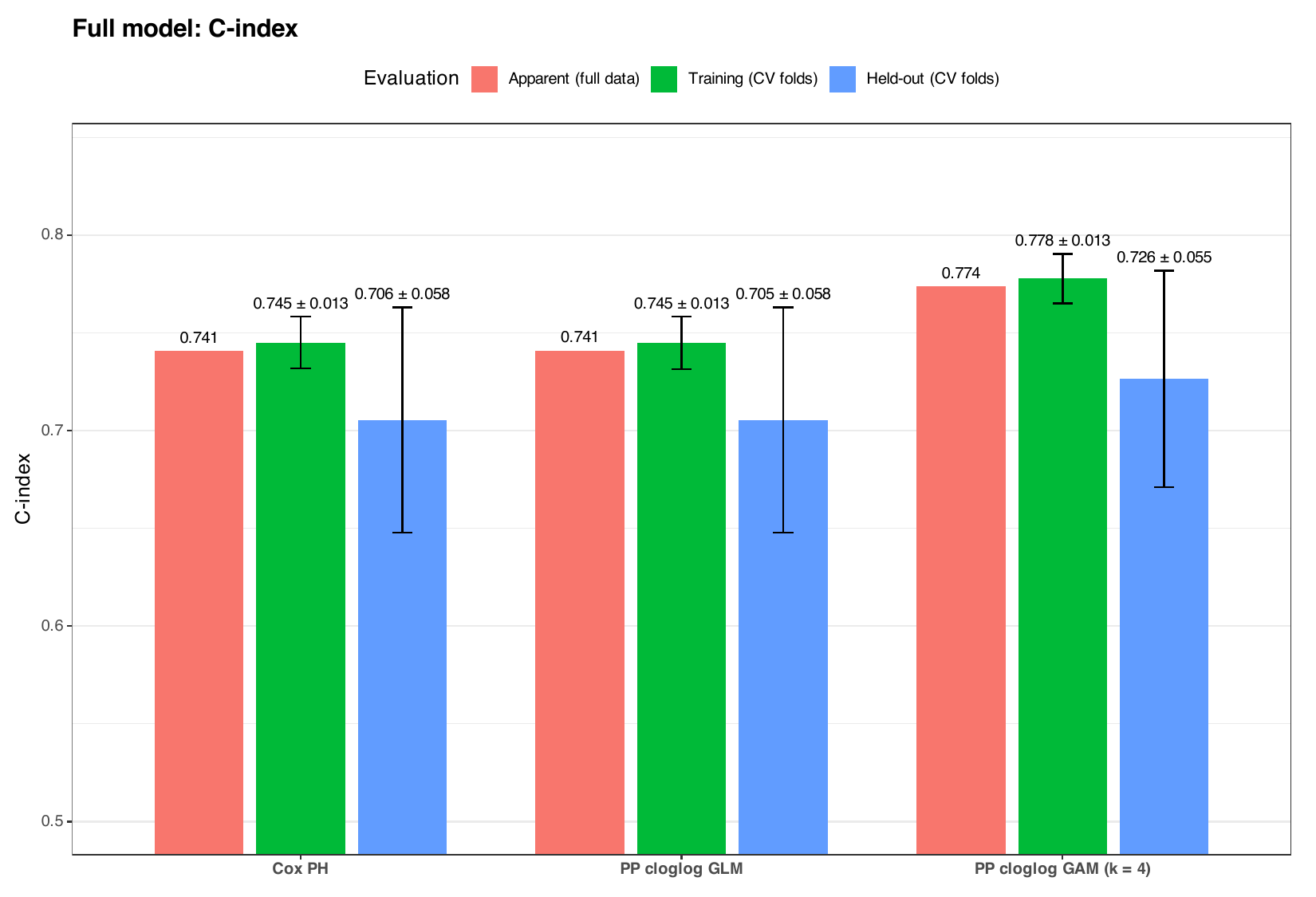}
        \caption{C-index}
        \label{fig:full_cindex}
    \end{subfigure}
    \hfill
    \begin{subfigure}[t]{0.32\textwidth}
        \centering
        \includegraphics[width=\linewidth]{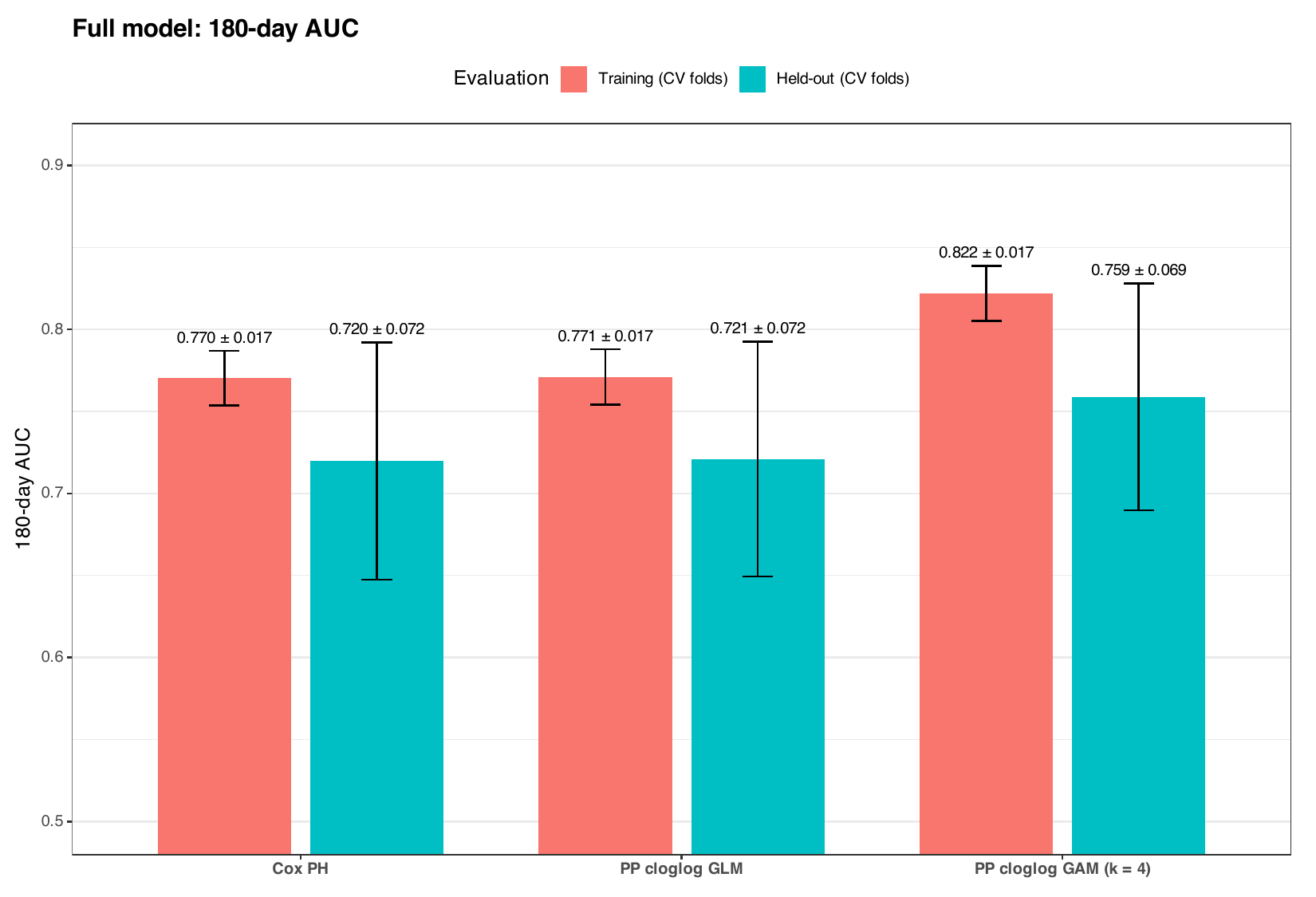}
        \caption{180-day time-dependent AUC}
        \label{fig:full_auc}
    \end{subfigure}
    \hfill
    \begin{subfigure}[t]{0.32\textwidth}
        \centering
        \includegraphics[width=\linewidth]{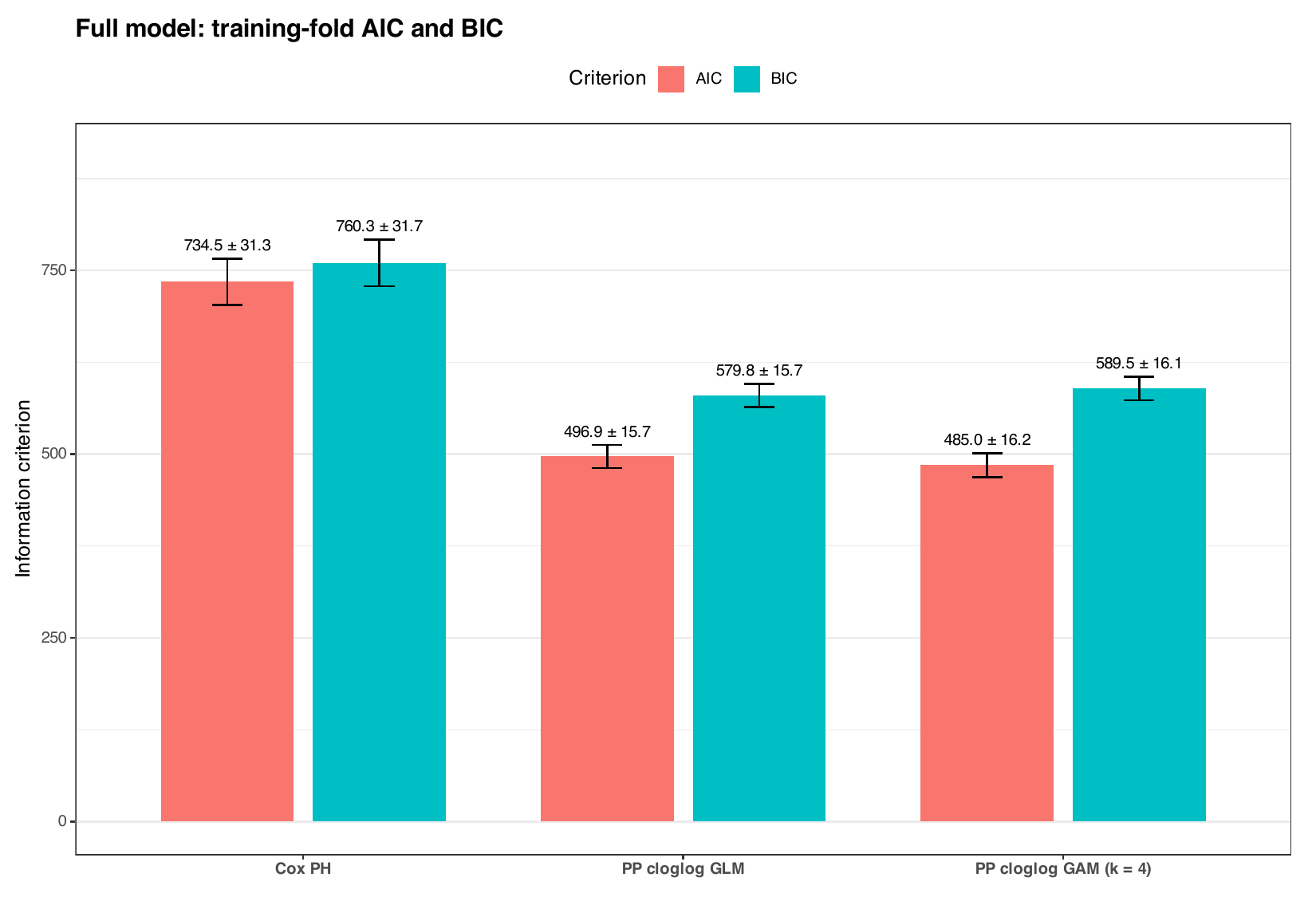}
        \caption{AIC and BIC}
        \label{fig:full_aic_bic}
    \end{subfigure}

    \caption{Repeated five-fold patient-level cross-validation results using
    all 11 baseline predictors. Error bars represent one standard deviation.
    AIC and BIC values should be compared only among models fitted using the
    same likelihood and data representation.}
    \label{fig:full_model_cv}
\end{figure*}

Figure~\ref{fig:full_model_cv} presents the complete cross-validation results,
including the means and standard deviations.

\subsubsection{Cross-validated performance using the feature-selected set}

A feature-selected set containing age, anaemia, creatinine phosphokinase,
ejection fraction, high blood pressure, and serum creatinine was evaluated
using the same repeated five-fold validation procedure. The reduced predictor
set produced modest improvements in held-out C-index and 180-day AUC across
all three statistical survival models.

Cox PH and the PP GLM remained nearly equivalent, while the PP GAM continued
to provide the strongest held-out discrimination. The GAM achieved a
held-out C-index of approximately 0.74 and a 180-day AUC of approximately
0.78. These results indicate that removing less informative predictors
slightly improved generalization without changing the relative ranking of the
models.

\begin{figure*}[htbp]
    \centering

    \begin{subfigure}[t]{0.32\textwidth}
        \centering
        \includegraphics[width=\linewidth]{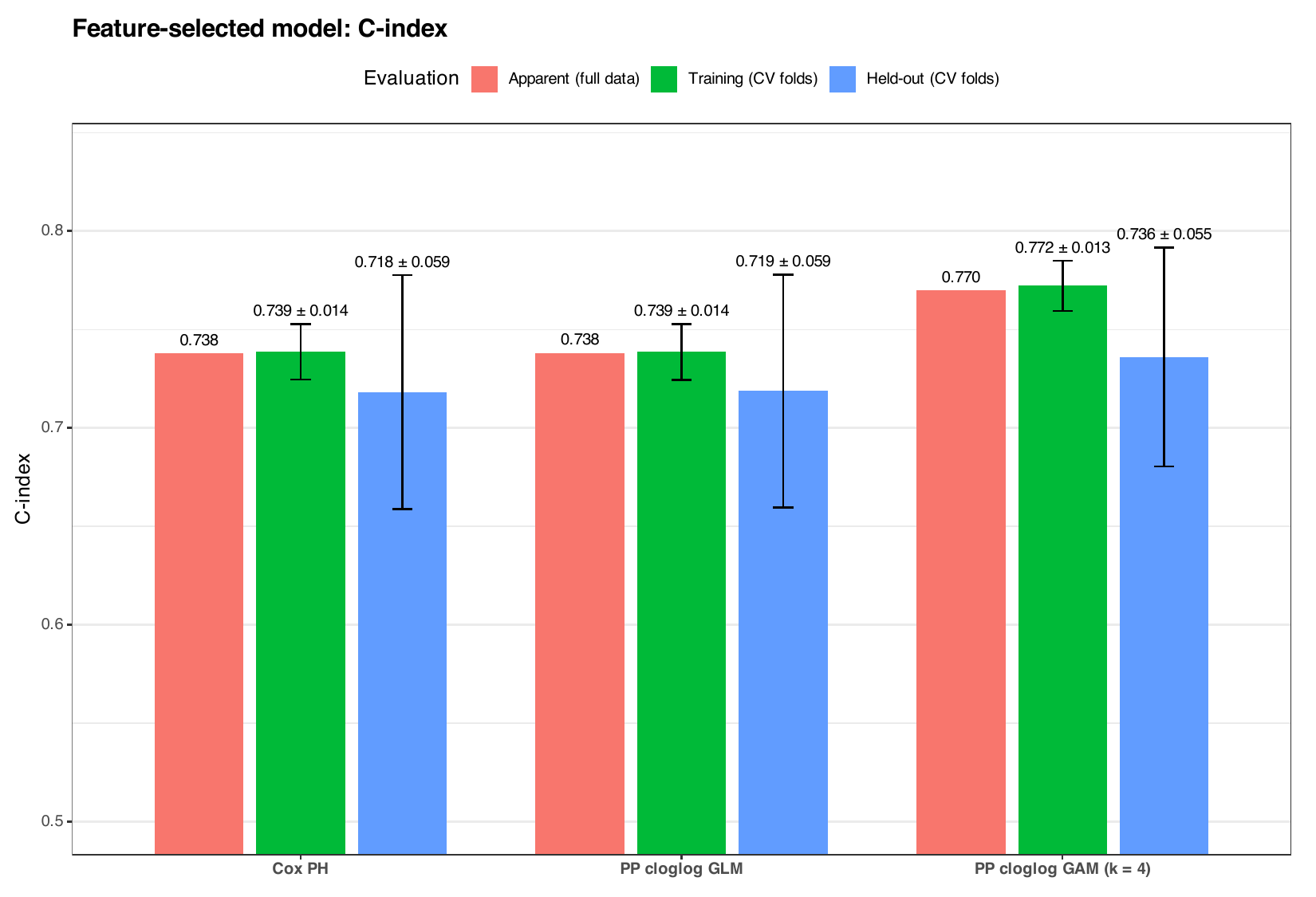}
        \caption{C-index}
        \label{fig:selected_cindex}
    \end{subfigure}
    \hfill
    \begin{subfigure}[t]{0.32\textwidth}
        \centering
        \includegraphics[width=\linewidth]{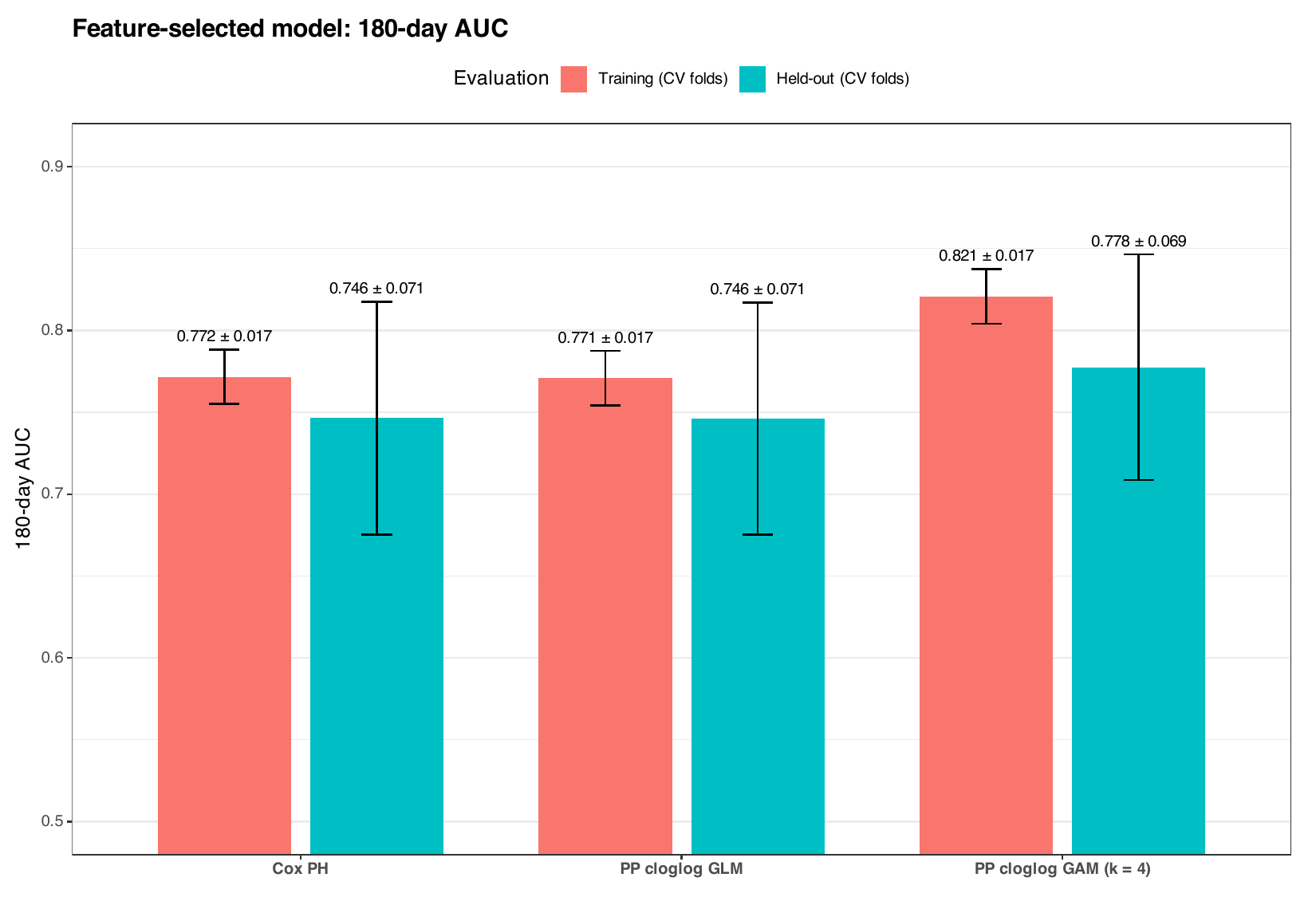}
        \caption{180-day time-dependent AUC}
        \label{fig:selected_auc}
    \end{subfigure}
    \hfill
    \begin{subfigure}[t]{0.32\textwidth}
        \centering
        \includegraphics[width=\linewidth]{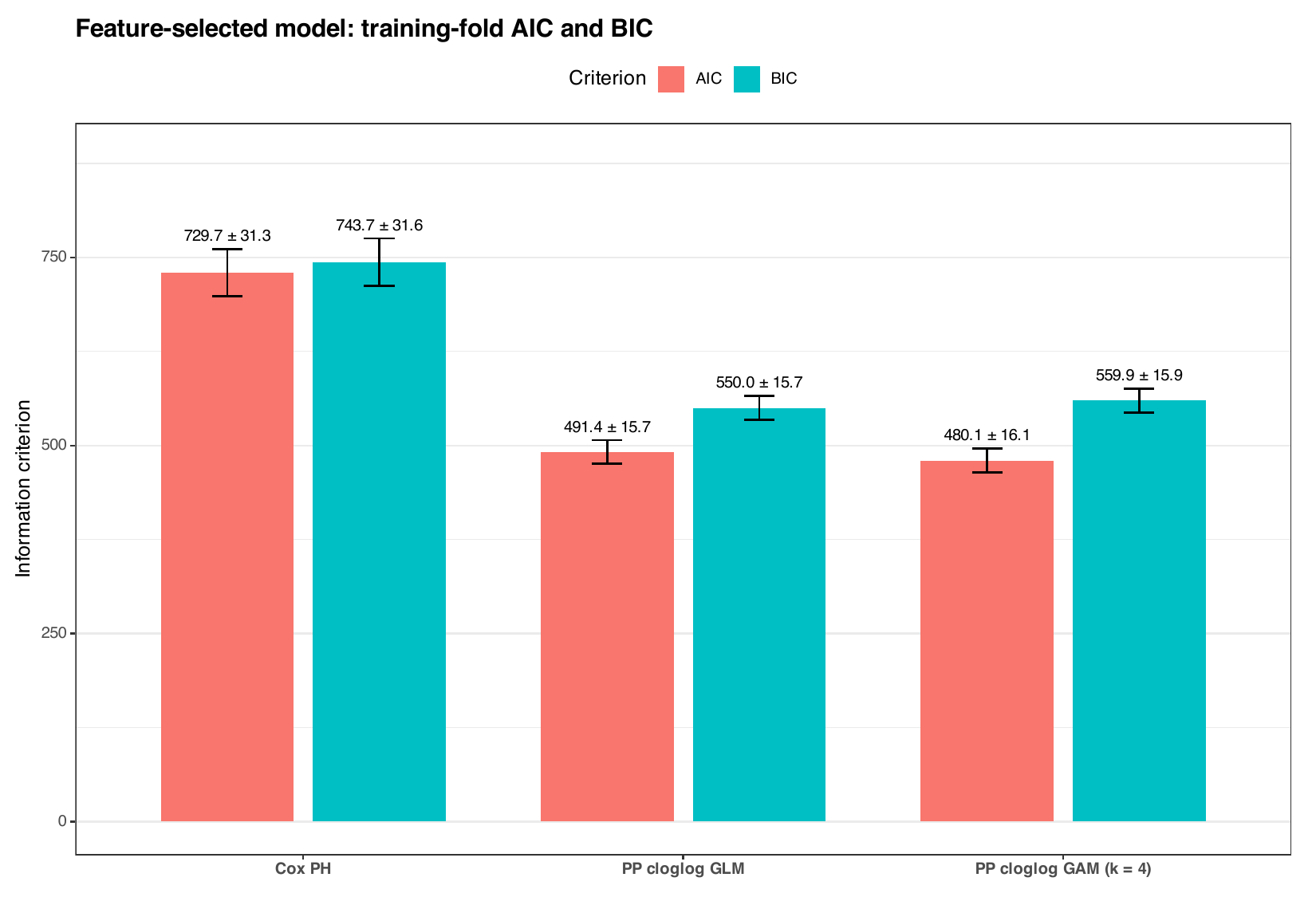}
        \caption{AIC and BIC}
        \label{fig:selected_aic_bic}
    \end{subfigure}

    \caption{Repeated five-fold cross-validation results using the
    feature-selected set of age, anaemia, creatinine phosphokinase, ejection
    fraction, high blood pressure, and serum creatinine. Error bars represent
    one standard deviation.}
    \label{fig:selected_model_cv}
\end{figure*}

AIC and BIC also decreased within each model family after feature selection,
indicating improved model simplicity. Among the feature-selected
person-period models, AIC continued to favor the PP GAM, whereas BIC favored
the simpler PP GLM. As in the full-predictor analysis, Cox PH information
criteria were interpreted only relative to other Cox specifications and were
not directly compared with those of the person-period models. Figure~\ref{fig:selected_model_cv} presents the complete feature-selected
results.

Overall, the PP GLM closely reproduced the predictive performance of Cox PH.
The PP GAM achieved the strongest held-out discrimination under both predictor
sets, while feature selection was associated with small improvements in
generalization and model simplicity.

\subsection{Predictive Performance and Generalization Across Survival Models}
\label{sec:model_performance_generalization}

The predictive performance comparison of statistical and machine learning survival models
was conducted using 100 repeated patient-level train--test splits. Performance
was evaluated on the held-out test sets, while the differences between
training and test performance were used to assess model generalization.

\subsubsection{Performance using the full predictor set}

Using all 11 baseline predictors, the PP random forest achieved the highest
mean held-out C-index, accuracy, and MCC. However, it also showed substantially
larger train--test gaps than Cox PH, the PP GLM, and the PP GAM. The PP
XGBoost model showed a similar pattern, with competitive held-out
classification performance but relatively large performance gaps. These
results suggest that the tree-based person-period models captured complex
patterns in the training data but were more susceptible to overfitting.

The PP GAM provided the highest mean 180-day AUC and PR AUC and the lowest
Brier score. Its train--test gaps were also considerably smaller than those
of the PP random forest and PP XGBoost. Therefore, although the PP random
forest achieved slightly higher C-index, accuracy, and MCC, the PP GAM
provided a better balance between held-out predictive performance and
generalization.

Cox PH and the PP GLM again produced similar held-out results and small
train--test gaps, further supporting the agreement between the continuous-
and discrete-time statistical formulations. The random survival forest
showed competitive discrimination but weaker accuracy and MCC. DeepSurv
produced the lowest held-out discrimination and the largest train--test gaps,
indicating the weakest generalization among the evaluated models.

\begin{table*}[http]
    \centering
    \caption{Held-out test performance of the survival models using all 11
    baseline predictors across 100 repeated patient-level train--test splits.
    Results are reported as mean $\pm$ one standard deviation.}
    \label{tab:full_model_test_performance}

    \footnotesize
    \setlength{\tabcolsep}{3pt}

    \begin{adjustbox}{max width=\textwidth}
    \begin{tabular}{lcccccc}
        \toprule
        \textbf{Model} &
        \textbf{C-index ($\uparrow$)} &
        \textbf{Brier Score ($\downarrow$)} &
        \textbf{AUC ($\uparrow$)} &
        \textbf{PR AUC ($\uparrow$)} &
        \textbf{Accuracy ($\uparrow$)} &
        \textbf{MCC ($\uparrow$)} \\
        \midrule

        Cox PH
        & $0.709 \pm 0.043$
        & $0.197 \pm 0.028$
        & $0.731 \pm 0.059$
        & $0.719 \pm 0.072$
        & $0.651 \pm 0.052$
        & $0.310 \pm 0.116$ \\

        PP GLM
        & $0.711 \pm 0.042$
        & $0.197 \pm 0.028$
        & $0.736 \pm 0.057$
        & $0.721 \pm 0.073$
        & $0.654 \pm 0.053$
        & $0.316 \pm 0.117$ \\

        PP GAM
        & $0.733 \pm 0.041$
        & $\mathbf{0.189 \pm 0.030}$
        & $\mathbf{0.774 \pm 0.057}$
        & $\mathbf{0.740 \pm 0.075}$
        & $0.689 \pm 0.047$
        & $0.391 \pm 0.101$ \\

        PP Random Forest
        & $\mathbf{0.734 \pm 0.041}$
        & $0.192 \pm 0.031$
        & $0.767 \pm 0.052$
        & $0.725 \pm 0.073$
        & $\mathbf{0.697 \pm 0.047}$
        & $\mathbf{0.398 \pm 0.096}$ \\

        PP XGBoost
        & $0.711 \pm 0.042$
        & $0.195 \pm 0.029$
        & $0.740 \pm 0.049$
        & $0.723 \pm 0.067$
        & $0.680 \pm 0.054$
        & $0.365 \pm 0.116$ \\

        Random Survival Forest
        & $0.732 \pm 0.046$
        & $0.198 \pm 0.014$
        & $0.763 \pm 0.053$
        & $0.737 \pm 0.068$
        & $0.591 \pm 0.041$
        & $0.206 \pm 0.104$ \\

        DNN (DeepSurv)
        & $0.642 \pm 0.049$
        & $0.279 \pm 0.046$
        & $0.656 \pm 0.062$
        & $0.647 \pm 0.072$
        & $0.626 \pm 0.054$
        & $0.247 \pm 0.113$ \\

        \bottomrule
    \end{tabular}
    \end{adjustbox}
\end{table*}

\begin{table*}[htbp]
    \centering
    \caption{Training-to-test performance gaps using all 11 baseline
    predictors across 100 repeated patient-level train--test splits.
    Smaller values indicate better generalization. Results are reported as
    mean $\pm$ one standard deviation.}
    \label{tab:full_model_performance_gaps}

    \small
    \setlength{\tabcolsep}{10pt}

    \begin{tabular}{lccc}
        \toprule
        \textbf{Model} &
        \textbf{C-index gap ($\downarrow$)} &
        \textbf{PR AUC gap ($\downarrow$)} &
        \textbf{MCC gap ($\downarrow$)} \\
        \midrule

        Cox PH
        & $0.037 \pm 0.060$
        & $0.036 \pm 0.101$
        & $0.061 \pm 0.159$ \\

        PP GLM
        & $0.035 \pm 0.059$
        & $0.037 \pm 0.103$
        & $0.064 \pm 0.160$ \\

        PP GAM
        & $0.049 \pm 0.055$
        & $0.070 \pm 0.100$
        & $0.075 \pm 0.139$ \\

        PP Random Forest
        & $\mathbf{0.187 \pm 0.045}$
        & $\mathbf{0.256 \pm 0.076}$
        & $\mathbf{0.395 \pm 0.114}$ \\

        PP XGBoost
        & $\mathbf{0.166 \pm 0.051}$
        & $\mathbf{0.210 \pm 0.077}$
        & $\mathbf{0.337 \pm 0.140}$ \\

        Random Survival Forest
        & $0.052 \pm 0.059$
        & $0.054 \pm 0.093$
        & $0.107 \pm 0.149$ \\

        DNN (DeepSurv)
        & $\mathbf{0.288 \pm 0.053}$
        & $\mathbf{0.324 \pm 0.074}$
        & $\mathbf{0.576 \pm 0.127}$ \\

        \bottomrule
    \end{tabular}
\end{table*}
\subsubsection{Performance using the feature-selected set}

The feature-selected set included age, anaemia, creatinine phosphokinase,
ejection fraction, high blood pressure, and serum creatinine. Compared with
the full predictor set, mean held-out performance improved across the
evaluated models. Train--test gaps also generally decreased, indicating
improved generalization with the reduced predictor set.

The PP random forest achieved the highest held-out C-index, AUC, accuracy,
and MCC. Its train--test gaps were smaller than those obtained with the full
predictor set, although they remained substantially larger than those of the
statistical models. Thus, feature selection reduced, but did not eliminate,
its tendency to overfit.

The PP GAM also showed strong held-out performance. It achieved the highest
PR AUC and tied with PP XGBoost for the lowest Brier score. In addition, its
train--test gaps were considerably smaller than those of the PP random forest
and PP XGBoost. These results indicate that the PP GAM provided the best
overall balance between predictive performance and generalization.

Feature selection improved the held-out performance of DeepSurv and reduced
its train--test gaps. However, DeepSurv continued to show the lowest C-index,
AUC, and PR AUC and relatively large performance gaps, indicating weaker
generalization than the statistical models. Cox PH, the PP GLM, and the
random survival forest maintained relatively small train--test gaps, nevertheless
their overall predictive performance remained below that of the PP GAM and
PP random forest.
\begin{table*}[htbp]
    \centering
    \caption{Held-out test performance of the survival models using the six
    feature-selected predictors across 100 repeated patient-level train--test
    splits. Results are reported as mean $\pm$ one standard deviation.}
    \label{tab:reduced_model_test_performance}

    \footnotesize
    \setlength{\tabcolsep}{3pt}

    \begin{adjustbox}{max width=\textwidth}
    \begin{tabular}{lcccccc}
        \toprule
        \textbf{Model} &
        \textbf{C-index ($\uparrow$)} &
        \textbf{Brier Score ($\downarrow$)} &
        \textbf{AUC ($\uparrow$)} &
        \textbf{PR AUC ($\uparrow$)} &
        \textbf{Accuracy ($\uparrow$)} &
        \textbf{MCC ($\uparrow$)} \\
        \midrule

        Cox PH
        & $0.718 \pm 0.045$
        & $0.187 \pm 0.023$
        & $0.751 \pm 0.057$
        & $0.741 \pm 0.070$
        & $0.670 \pm 0.050$
        & $0.352 \pm 0.112$ \\

        PP GLM
        & $0.721 \pm 0.043$
        & $0.187 \pm 0.025$
        & $0.755 \pm 0.056$
        & $0.740 \pm 0.072$
        & $0.674 \pm 0.045$
        & $0.359 \pm 0.098$ \\

        PP GAM
        & $0.741 \pm 0.040$
        & $\mathbf{0.182 \pm 0.027}$
        & $0.789 \pm 0.055$
        & $\mathbf{0.754 \pm 0.073}$
        & $0.702 \pm 0.048$
        & $0.420 \pm 0.097$ \\

        PP Random Forest
        & $\mathbf{0.748 \pm 0.042}$
        & $0.185 \pm 0.029$
        & $\mathbf{0.793 \pm 0.052}$
        & $0.744 \pm 0.073$
        & $\mathbf{0.712 \pm 0.044}$
        & $\mathbf{0.433 \pm 0.088}$ \\

        PP XGBoost
        & $0.732 \pm 0.042$
        & $\mathbf{0.182 \pm 0.028}$
        & $0.774 \pm 0.053$
        & $0.751 \pm 0.068$
        & $0.709 \pm 0.052$
        & $0.425 \pm 0.108$ \\

        Random Survival Forest
        & $0.738 \pm 0.046$
        & $0.193 \pm 0.015$
        & $0.777 \pm 0.054$
        & $0.748 \pm 0.071$
        & $0.618 \pm 0.043$
        & $0.259 \pm 0.093$ \\

        DNN (DeepSurv)
        & $0.687 \pm 0.045$
        & $0.218 \pm 0.036$
        & $0.727 \pm 0.059$
        & $0.696 \pm 0.076$
        & $0.659 \pm 0.052$
        & $0.316 \pm 0.111$ \\

        \bottomrule
    \end{tabular}
    \end{adjustbox}
\end{table*}

\begin{table*}[htbp]
    \centering
    \caption{Training-to-test performance gaps using the six feature-selected
    predictors across 100 repeated patient-level train--test splits. Smaller
    values indicate better generalization. Results are reported as mean
    $\pm$ one standard deviation.}
    \label{tab:reduced_model_performance_gaps}

    \small
    \setlength{\tabcolsep}{10pt}

    \begin{tabular}{lccc}
        \toprule
        \textbf{Model} &
        \textbf{C-index gap ($\downarrow$)} &
        \textbf{PR AUC gap ($\downarrow$)} &
        \textbf{MCC gap ($\downarrow$)} \\
        \midrule

        Cox PH
        & $0.022 \pm 0.063$
        & $\mathbf{0.017 \pm 0.099}$
        & $\mathbf{0.040 \pm 0.156}$ \\

        PP GLM
        & $\mathbf{0.019 \pm 0.061}$
        & $\mathbf{0.017 \pm 0.102}$
        & $0.045 \pm 0.147$ \\

        PP GAM
        & $0.035 \pm 0.054$
        & $0.050 \pm 0.097$
        & $0.053 \pm 0.132$ \\

        PP Random Forest
        & $0.150 \pm 0.049$
        & $0.216 \pm 0.080$
        & $0.291 \pm 0.111$ \\

        PP XGBoost
        & $0.119 \pm 0.053$
        & $0.157 \pm 0.083$
        & $0.227 \pm 0.137$ \\

        Random Survival Forest
        & $0.041 \pm 0.059$
        & $0.039 \pm 0.098$
        & $0.092 \pm 0.134$ \\

        DNN (DeepSurv)
        & $0.162 \pm 0.056$
        & $0.206 \pm 0.090$
        & $0.304 \pm 0.143$ \\

        \bottomrule
    \end{tabular}
\end{table*}

Across both predictor sets, no single model was superior according to every
performance measure. The PP random forest generally achieved the strongest
held-out C-index and classification performance, but its relatively large
train--test gaps indicated overfitting. In contrast, the PP GAM
combined strong discrimination, favorable Brier and PR AUC results, and
smaller performance gaps.

The PP GLM remained closely aligned with Cox PH, confirming that the
person-period representation preserved the predictive behavior of the
continuous-time statistical model. The PP XGBoost model provided competitive
held-out performance but showed greater overfitting, while DeepSurv performed
poorly relative to the other approaches.

Overall, reducing the predictor set was associated with higher mean held-out
performance and smaller train--test gaps across all models. The PP GAM provided the strongest overall balance between predictive accuracy and generalization, whereas the PP random forest achieved the highest performance on several individual held-out metrics.

\subsection{Effect of Including Observed Follow-Up Duration}
\label{sec:followup_effect}

To evaluate the effect of observed follow-up duration on fixed-horizon
mortality prediction, three predictor scenarios were compared, clinical
variables only, clinical variables combined with follow-up duration, and
follow-up duration only. Logistic regression, weighted logistic regression,
random forest, and XGBoost were evaluated using 100 repeated patient-level
train--test splits.
\begin{table*}[htbp]
    \centering
    \caption{Mortality prediction performance with and without observed
    follow-up duration across 100 repeated patient-level train--test splits.
    Results are reported as mean $\pm$ one standard deviation.}
    \label{tab:followup_comparison}

    \footnotesize
    \setlength{\tabcolsep}{4pt}

    \begin{adjustbox}{max width=\textwidth}
    \begin{tabular}{llccccc}
        \toprule
        \textbf{Model} &
        \textbf{Feature scenario} &
        \textbf{AUC ($\uparrow$)} &
        \textbf{PR AUC ($\uparrow$)} &
        \textbf{Brier score ($\downarrow$)} &
        \textbf{MCC ($\uparrow$)} &
        \textbf{Accuracy ($\uparrow$)} \\
        \midrule

        Logistic
        & Clinical only
        & $0.7268 \pm 0.0642$
        & $0.7064 \pm 0.0672$
        & $0.2150 \pm 0.0277$
        & $0.3476 \pm 0.1175$
        & $0.6764 \pm 0.0577$ \\

        Logistic
        & Clinical + follow-up
        & $0.9741 \pm 0.0339$
        & $0.9916 \pm 0.0149$
        & $0.0311 \pm 0.0318$
        & $0.9386 \pm 0.0643$
        & $0.9689 \pm 0.0325$ \\

        Logistic
        & Follow-up only
        & $0.9890 \pm 0.0117$
        & $0.9998 \pm 0.0004$
        & $0.0100 \pm 0.0105$
        & $0.9802 \pm 0.0213$
        & $0.9900 \pm 0.0107$ \\
        \midrule

        Weighted logistic
        & Clinical only
        & $0.7274 \pm 0.0642$
        & $0.7070 \pm 0.0674$
        & $0.2159 \pm 0.0281$
        & $0.3601 \pm 0.1190$
        & $0.6805 \pm 0.0591$ \\

        Weighted logistic
        & Clinical + follow-up
        & $0.9738 \pm 0.0344$
        & $0.9916 \pm 0.0149$
        & $0.0316 \pm 0.0315$
        & $0.9379 \pm 0.0643$
        & $0.9686 \pm 0.0325$ \\

        Weighted logistic
        & Follow-up only
        & $0.9890 \pm 0.0117$
        & $0.9998 \pm 0.0004$
        & $0.0100 \pm 0.0104$
        & $0.9802 \pm 0.0213$
        & $0.9900 \pm 0.0107$ \\
        \midrule

        Random forest
        & Clinical only
        & $0.7525 \pm 0.0506$
        & $0.6946 \pm 0.0628$
        & $0.2042 \pm 0.0221$
        & $0.3934 \pm 0.1054$
        & $0.6981 \pm 0.0527$ \\

        Random forest
        & Clinical + follow-up
        & $0.9953 \pm 0.0086$
        & $0.9955 \pm 0.0075$
        & $0.0241 \pm 0.0122$
        & $0.9665 \pm 0.0372$
        & $0.9828 \pm 0.0193$ \\

        Random forest
        & Follow-up only
        & $0.9872 \pm 0.0161$
        & $0.9998 \pm 0.0004$
        & $0.0123 \pm 0.0150$
        & $0.9769 \pm 0.0286$
        & $0.9883 \pm 0.0147$ \\
        \midrule

        XGBoost
        & Clinical only
        & $0.7417 \pm 0.0522$
        & $0.7151 \pm 0.0636$
        & $0.2074 \pm 0.0231$
        & $0.3991 \pm 0.1059$
        & $0.7001 \pm 0.0530$ \\

        XGBoost
        & Clinical + follow-up
        & $0.9959 \pm 0.0057$
        & $0.9963 \pm 0.0048$
        & $0.0113 \pm 0.0092$
        & $0.9802 \pm 0.0213$
        & $0.9900 \pm 0.0107$ \\

        XGBoost
        & Follow-up only
        & $0.9890 \pm 0.0117$
        & $0.9998 \pm 0.0004$
        & $0.0108 \pm 0.0097$
        & $0.9802 \pm 0.0213$
        & $0.9900 \pm 0.0107$ \\

        \bottomrule
    \end{tabular}
    \end{adjustbox}
\end{table*}
First, using clinical variables alone resulted in moderate predictive performance
across all four models. Mean AUC values ranged from approximately 0.73 to
0.75, and mean accuracy ranged from approximately 0.68 to 0.70. Random forest
achieved the highest clinical-only AUC, although its performance was only
slightly higher than that of the other models.

Second, adding observed follow-up duration to baseline predictors resulted in a significant increase in all
performance measures. The mean AUC increased to approximately 0.97--1.00
with accuracy increased to approximately 0.97--0.99. MCC values also
increased from approximately 0.35--0.40 with clinical variables alone to
approximately 0.94--0.98 after follow-up duration was added. Additionally,
the Brier scores decreased, indicating more accurate
predicted probabilities. This pattern was consistent across the statistical
and machine-learning classifiers.

Third, follow-up duration alone also produced near-perfect classification. Across
the four models, its mean AUC ranged from approximately 0.987 to 0.989 with
the mean PR AUC was approximately 1.00. Accuracy was approximately 0.99, and
MCC values were approximately 0.98. The strong performance obtained without
any clinical predictors indicates that most of the improvement from
the combined predictor set was driven by the observed follow-up duration
rather than by additional clinical information.

Observed follow-up duration is determined after baseline and is directly
related to both event occurrence and censoring. In particular, death
terminates follow-up, while patients who remain alive may contribute longer
observation periods. Consequently, including observed follow-up duration as
a baseline predictor provides the classifier with information closely
connected to the outcome. The resulting near perfect performance
therefore does not represent clinically meaningful prediction.

These findings distinguish the appropriate and inappropriate uses of
follow-up duration. In survival analysis, follow-up duration is required to
define event times and censoring and is therefore part of the outcome
structure. However, it should not be included as a baseline predictor when
estimating mortality risk at the beginning of follow-up. Accordingly,
observed follow-up duration was excluded from the predictor sets used in all
primary survival-model comparisons.

Overall, the analysis demonstrates that the apparent near-perfect prediction
obtained after including follow-up duration was primarily driven by
post-baseline outcome-related information rather than by improved learning
from the clinical predictors.

\section{Conclusion and Future Work}
\label{sec:conclusion}

This study developed a survival-aware framework for heart-failure mortality
prediction that preserves the time-to-event structure of the data. A
discrete-time person-period transformation recast the survival outcome as an
interval-level binary problem retaining event timing and right-censoring while
allowing conventional classifiers to be applied. 

The person-period GLM reproduced the hazard ratios and concordance of the Cox
proportional hazards model validating the transformation. The person-period
GAM captured significant nonlinear effects for age, ejection fraction, and
serum creatinine and gave the best balance of discrimination and
generalization. The person-period random forest achieved the strongest raw
classification performance but overfit substantially while DeepSurv
generalized worst. Feature selection improved held-out performance across all
models without changing their ordering. Finally, including observed follow-up
duration inflated classification performance to near-perfect levels,
confirming that it must not be used as a baseline predictor.

Future work should validate the framework using larger and independent patient
cohorts, as the present study was based on a single dataset of 299 patients.
Additional studies should examine alternative interval structures and investigate
time-varying predictor effects. External validation will also be necessary
before the proposed framework can be considered for clinical risk assessment.

\printbibliography[title={References}]

\end{document}